\documentclass[apj]{emulateapj}
\usepackage{graphics}
\usepackage{natbib}
\newcommand{\as}{$''$}
\newcommand{\am}{$'$}
\newcommand{\nh}{$N_{\rm{H}}$}
\newcommand{\nustar}{\textit{NuSTAR}}
\newcommand{\pksb}{PKS\,2155-304}
\newcommand{\qcs}{3C\,273}

\begin{document}

\title{Calibration of the \nustar\ High Energy Focusing X-ray Telescope.}
\author{Kristin K. Madsen$^1$, Fiona A. Harrison$^1$, Craig B. Markwardt$^2$, Hongjun An$^3$, Brian W. Grefenstette$^1$, Matteo Bachetti$^{11,12,13}$, Hiromasa Miyasaka$^1$, Takao Kitaguchi$^4$, Varun Bhalerao$^5$, Steve Boggs$^{6}$, Finn E. Christensen$^7$, William W. Craig$^6$, Karl Forster$^1$, Felix Fuerst$^1$, Charles J. Hailey$^8$, Matteo Perri$^9$, Simonetta Puccetti$^9$, Vikram Rana$^1$, Daniel Stern$^{10}$, Dominic J. Walton$^{10,1}$, Niels J\o rgen Westergaard$^7$, and William W. Zhang$^2$}

\affiliation{
$^1$ Cahill Center for Astronomy and Astrophysics, California Institute of Technology, Pasadena, CA 91125, USA\\
$^2$ Goddard Space Flight Center, Greenbelt, MD 20771, USA\\
$^3$ Department of Physics, McGill University, Montreal, Quebec, H3A 2T8, Canada\\
$^4$ RIKEN, 2-1 Hirosawa, Wako, Saitama, 351-0198, Japan\\
$^5$ Inter-University Center for Astronomy and Astrophysics, Post Bag 4, Ganeshkhind, Pune 411007, India\\
$^6$ Space Sciences Laboratory, University of California, Berkeley, CA 94720, USA\\
$^7$ DTU Space, National Space Institute, Technical University of Denmark, Elektronvej 327, DK-2800 Lyngby, Denmark\\
$^8$ Columbia Astrophysics Laboratory, Columbia University, New York 10027, USA\\
$^9$ ASI Science Data Center, via Galileo Galilei, I-00044, Frascati, Italy\\
$^{10}$ Jet Propulsion Laboratory, California Institute of Technology, Pasadena, CA 91109, USA\\
$^{11}$ Universit\'e de Toulouse; UPS-OMP; IRAP; Toulouse, France\\
$^{12}$ CNRS; Institut de Recherche en Astrophysique et Plan\'etologie; 9 Av. colonel Roche, BP 44346, F-31028 Toulouse cedex 4, France\\
$^{13}$Osservatorio Astronomico di Cagliari\\
}

\begin{abstract}
We present the calibration of the \textit{Nuclear Spectroscopic Telescope Array} (\nustar) X-ray satellite. We used the Crab as the primary effective area calibrator and constructed a piece-wise linear spline function to modify the vignetting response. The achieved residuals for all off-axis angles and energies, compared to the assumed spectrum, are typically better than $\pm 2$\% up to 40\,keV and 5--10\,\% above due to limited counting statistics. An empirical adjustment to the theoretical 2D point spread function (PSF) was found using several strong point sources, and no increase of the PSF half power diameter (HPD) has been observed since the beginning of the mission. We report on the detector gain calibration, good to 60\,eV for all grades, and discuss the timing capabilities of the observatory, which has an absolute timing of $\pm$ 3\,ms. Finally we present cross-calibration results from two campaigns between all the major concurrent X-ray observatories (\textit{Chandra}, \textit{Swift}, \textit{Suzaku} and \textit{XMM-Newton}), conducted in 2012 and 2013 on the sources 3C\,273 and PKS\,2155-304, and show that the differences in measured flux is within $\sim$10\% for all instruments with respect to \nustar.
\end{abstract}

\keywords{space vehicles: instruments -- X-rays: individual (3C\,273) -- X-rays: individual (PKS\,2155-304)}

\section{Introduction}
The \textit{Nuclear Spectroscopic Telescope Array} (\nustar) was successfully launched in June 2012 \citep{Harrison2013}. \nustar\ carries two co-aligned conical Wolter-I approximation \citep{Petre1985} Optics Modules (OMA, and OMB) that focus onto two identical Focal Plane Modules modules (FPMA and FPMB), each composed of four solid state CdZnTe pixel detector arrays (enumerated DET0 through DET3) with a minimum detector threshold of 3\,keV \citep{Rana2009,Kitaguchi2011}. There are 133 shells in each optic. The outer 43 shells are coated with a W/Si multilayer while the inner 90 shells are coated with Pt/C, limiting the highest efficient reflective X-ray energies below the Pt 78.4 keV K-edge \citep{Madsen2009}. The field of view is 13\am$\times$13\am\ and the physical pixel size is 12.3\as. Sub pixel resolution is obtained for events sharing charge among multiple pixels, and the physical pixel is subdivided by a factor of five in software to an effective pixel size of 2.5\as.

Extensive ground calibration and modeling of the subsystems were performed prior to launch \citep{Rana2009,Koglin2011,Brejnholt2011,Kitaguchi2011,Brejnholt2012,Westergaard2012}, and in this paper we describe the in-orbit refinement of the calibration of the \textit{NuSTAR} observatory. In \S\,\ref{effarea} we present the in-orbit effective area calibration, detailing the various components that go into generating the correct responses; in \S\,\ref{psf} we discuss the in-orbit point spread function (PSF) calibration; in \S\,\ref{gainsection} we show detector gain calibration, and in \S\,\ref{crosscalibration} we preset the cross-calibration between \nustar\ and the concurrent observatories: \textit{Chandra} \citep{Weisskopf2002}, \textit{INTEGRAL} \citep{Winkler2003}, \textit{NuSTAR} \citep{Harrison2013}, \textit{Suzaku} \citep{Mitsuda2007}, \textit{Swift} \citep{Gehrels2004}, and \textit{XMM-Newton} \citep{Jansen2001}. Finally \S\,\ref{time} discusses the timing calibration. 

\begin{table}
\caption{\textit{NuSTAR} Observations log of the Crab}
\centering
\begin{tabular}{cccc}
\hline
Obs ID & UT Date & \footnote{Effective exposure time corrected for dead-time.}Exposure & Off-axis angle A/B \\
& start & (seconds) & (arcminutes) \\
\hline 
\hline
10013022002 & 2012-09-20 & 2592 &  1.52/ 2.05 \\
10013022004 & 2012-09-21 & 2347 &  1.51/ 2.03 \\
10013022006 & 2012-09-21 & 2587 &  1.49/ 2.02 \\
10013023002 & 2012-09-21 & 2102 &  4.01/ 3.38 \\
10013024002 & 2012-09-20 & 2258 &  4.77/ 4.20 \\
10013025002 & 2012-09-25 & 1235 &  5.08/ 5.68 \\
10013025004 & 2012-09-25 & 1161 &  5.02/ 5.62 \\
10013025006 & 2012-09-27 & 1593 &  4.95/ 5.55 \\
10013026002 & 2012-09-26 & 2540 &  5.56/ 5.64 \\
10013026004 & 2012-09-26 & 1162 &  5.57/ 5.65 \\
10013027002 & 2012-09-27 & 1182 &  5.43/ 4.99 \\
10013027004 & 2012-09-27 & 1105 &  5.48/ 5.03 \\
10013028002 & 2012-09-28 & 1601 &  6.30/ 5.82 \\
10013028004 & 2012-09-28 & 1254 &  6.31/ 5.82 \\
10013029001 & 2012-10-25 & 2909 &  4.07/ 3.41 \\
10013030001 & 2012-10-25 & 3023 &  4.65/ 5.29 \\
10013031002 & 2012-10-25 & 2507 &  2.19/ 2.86 \\
10013032002 & 2012-11-04 & 2595 &  1.09/ 1.27 \\
10013033002 & 2012-12-19 & 1383 &  1.30/ 1.93 \\
10013033004 & 2012-12-21 & 1269 &  1.19/ 0.61 \\
10013034002 & 2013-02-14 & 988 &  1.02/ 1.69 \\
10013034004 & 2013-02-14 & 5720 &  0.54/ 0.88 \\
10013034005 & 2013-02-15 & 5968 &  0.75/ 0.87 \\
10013035002 & 2013-02-15 & 9401 &  5.32/ 5.44 \\
10013037002 & 2013-04-03 & 2679 &  2.14/ 2.65 \\
10013037004 & 2013-04-04 & 2796 &  3.28/ 3.95 \\
10013037006 & 2013-04-05 & 2799 &  3.24/ 3.90 \\
10013037008 & 2013-04-18 & 2814 &  3.28/ 3.95 \\
10013038002 & 2013-04-08 & 3084 &  4.20/ 4.78 \\
10013038004 & 2013-04-09 & 2217 &  4.21/ 4.79 \\
10013038006 & 2013-04-10 & 266 &  4.42/ 5.01 \\
10013038008 & 2013-04-17 & 2231 &  4.16/ 4.74 \\
10013039001 & 2013-05-01 & 5206 &  3.79/ 3.39 \\
10013039002 & 2013-05-02 & 590 &  3.66/ 3.25 \\
10013039003 & 2013-05-03 & 583 &  3.66/ 3.25 \\
80001022002 & 2013-03-09 & 3917 &  1.42/ 1.75 \\
10002001002 & 2013-09-02 & 2608 &  1.72/ 2.09 \\
10002001004 & 2013-09-03 & 2386 &  1.73/ 2.26 \\
10002001006 & 2013-11-11 & 14260 &  1.08/ 1.28 \\
\hline
\end{tabular}
\label{calobsid}
\end{table} 

\section{Effective area calibation}\label{effarea}
\subsection{The Crab as calibration source}
The Crab is a center filled pulsar wind nebula (PWN) powered by a pulsar with a double peaked profile of period P $\sim$ 33\,ms. It has served as the primary celestial calibration source for many hard X-ray instruments because of its brightness, relative stability, and simple power-law spectrum over the band from 1--100\,keV \citep{Kirsch2005}. The Crab, however, is too bright for reliable pileup corrections for most CCD based focusing X-ray instruments, and has been replaced with fainter sources, such as the Crab-like PWN G21.5+0.9 \citep{Tsujimoto2011}, the blazar PKS\,2155-304 \citep{Ishida2011} or the quasar 3C\,273. The latter two are variable and only serve for cross-calibration among observatories, while G21.5+0.9 would require unfeasible long integration times for \nustar\ to obtain the required statistics to calibrate the response at 80\,keV. The Crab therefore remains the best choice for the internal calibration in the X-ray band covered by \textit{NuSTAR}.

The spatially phase-averaged integrated spectrum of the Crab nebula+pulsar in the 1--100\,keV X-ray band has been well-described by a power-law with photon index of $\Gamma\sim 2.1$ \citep[\textit{RXTE, BeppoSAX, EXOSAT, INTEGRAL}/JEM\_X;][]{Kirsch2005}. Above 100\,keV the hard X-ray instruments (\textit{INTEGRAL}/SPI/ISGRI, \textit{CGRO}) measure a softer index of $\Gamma \sim $2.20--2.25, and below 10\,keV the instruments with CCD detectors a harder spectrum ($\Gamma < 2.1$). The softening above 100\,keV primarily comes from the curvature of the pulsed spectrum \citep{Kuiper2001}. The hardening in CCD X-ray instruments may be due to photon pile-up and although models exist for these instruments to deal with the pile-up the Crab usually still challenges the pile-up corrections and requires special non-standard reductions. \citet{Weisskopf2010} did a study with \textit{RXTE}/PCA, \textit{XMM}, and \textit{ASCA} looking for deviations from a power-law in the 0.2--50\,keV energy range and concluded that within the precision of the available instrumentation there is no detectable bend in the phase-averaged integrated spectrum of the Crab. Over the 16 years {\em RXTE } was operational and regularly monitoring the Crab, the spectral index was seen to vary by a peak-to-peak variation of $\Delta \Gamma \sim 0.025$ \citep{Shaposhnikov2012}. This variation is consistent with the observed spread between instruments, but it is slow and on average over the 16 years the Crab has remained at $\Gamma = 2.1$. 

Because the non piled-up instruments covering the 1--100\,keV band agree on a photon index of $\Gamma = 2.1 \pm 0.02$ and do not measure any curvature in the spectrum across this band, we have calibrated the effective area against a Crab index of $\Gamma = 2.1$. We model the differential photon spectral model of the phase-averaged integrated spectrum of pulsar+nebula as a simple absorbed power law,
\begin{eqnarray}
{dN(E)\over dE} = {\rm tbabs}(E) N E^{-\Gamma} {\rm ph}\,{\rm s}^{-1} {\rm cm}^{-2} {\rm keV}^{-1}
\label{crabspectmodel}
\end{eqnarray}
Here $E$ is the photon energy, \texttt{tbabs} is the interstellar absorption using Wilms abundances \citep{Wilms2000} and Verner cross-sections \citep{Verner1996}, $\Gamma$ is the power-law photon index, and $N$ is the normalization factor. When $\Gamma = 2.1$ and N = 10 keV$^{-1}$ cm$^{-2}$ s$^{-1}$ at 1 keV, we refer to this as the canonical Crab model.

In terms of flux stability the Crab has been tracked in great detail over the last decade. It remains largely stable, albeit with flux changes as much as 7\% \citep{Wilson2011} across the 10--100\,keV bandpass. This corresponds to a decay in the flux of 3.5\% per year over the period it has been observed. The \textit{NuSTAR} observations span approximately a year, but due to the optical axis knowledge (to be discussed in \S\,\ref{repeatability}) we expect flux differences between repeated observations to be on the same order (5\%), and therefore the intrinsic variability of the Crab is not an issue. In addition, after adjustment of the effective areas of the two mirror modules assuming the canonical Crab model, we adjust the overall normalization by 15\% and use the two targets PKS\,2155-304 and 3C\,273 observed simultaneously with \textit{Chandra}, \textit{NuSTAR}, \textit{Swift}, \textit{Suzaku} and \textit{XMM-Newton} to confirm the final global normalization of our responses with respect to the other observatories.

\begin{figure}
\begin{center}
\includegraphics[width=0.47\textwidth]{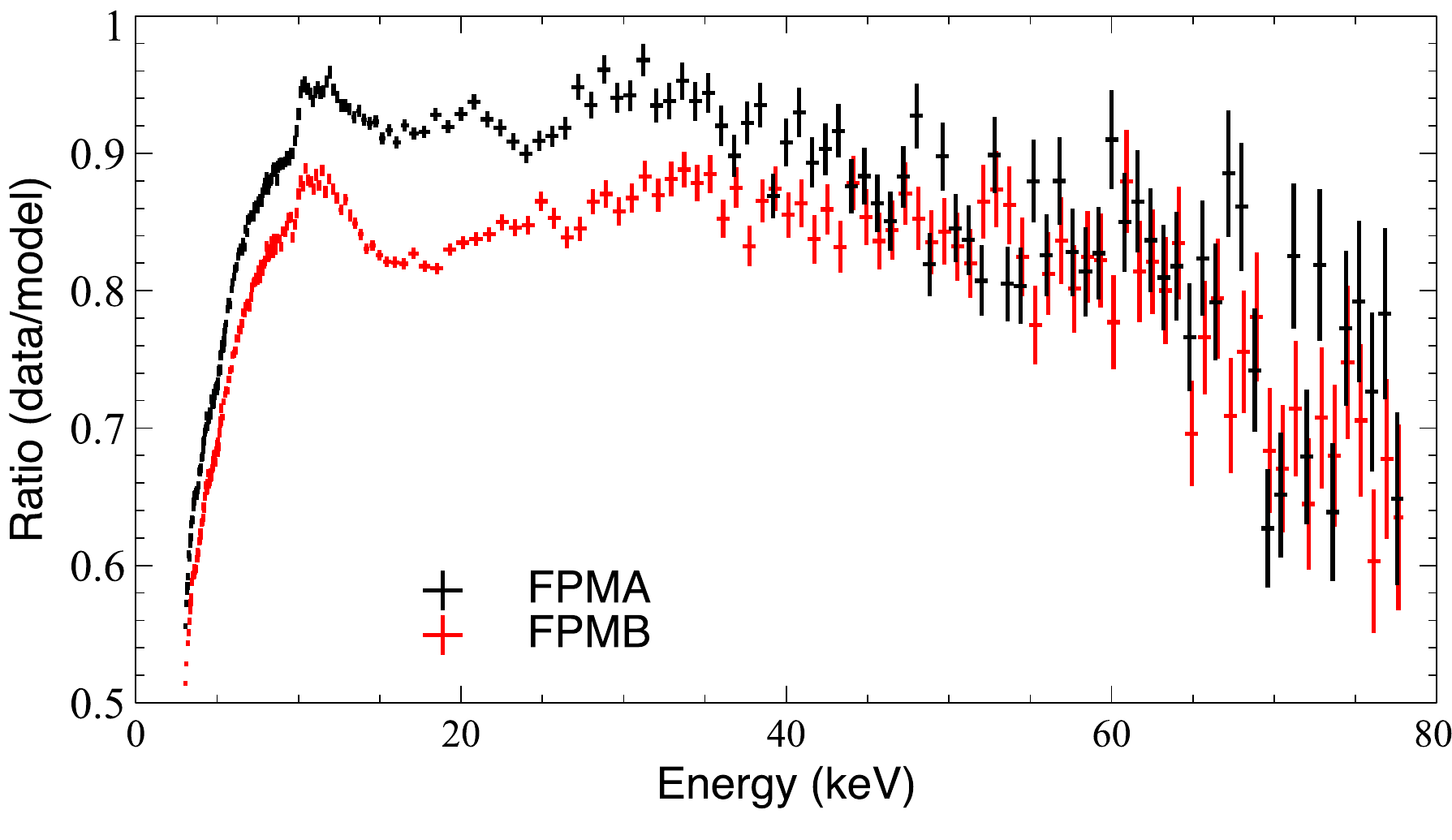}
\end{center}
\caption{Ratio of the base effective area to model ($\Gamma$=2.1, N=10 ph keV$^{-1}$ cm$^{-2}$ s$^{-1}$) for observation 10002001006, which was the longest single continuous observation taken at a fixed off-axis angle.}
\label{beforecrab}
\end{figure}

\begin{table}[tr]
\centering
\caption{Edges}
\begin{tabular}{c|c|c}
\hline
Edge (keV)& W & Pt \\
\hline
\hline
$L_3$ & 10.21 & 11.56 \\
$L_2$ & 11.54 & 13.27\\
$L_1$ & 12.10 & 13.88 \\
$K$ & 69.53 & 78.40 \\
\hline
\end{tabular}
\label{edges}
\end{table}

\subsection{Observations and Data Reduction}
The \textit{NuSTAR} Crab calibration campaign measured the Crab spectrum at different locations on the detector modules at various off-axis angles. This resulted in the 39 individual observations listed in Table \ref{calobsid}. 

Despite the angular extent of the Crab of $\sim$120\as$\times100$\as, we treat it as a point source since the center of the nebula where the pulsar resides dominates the emission. The difference in the effective area response between using a point source effective area centered on the pulsar and a count-rate adjusted extended effective area is only $\sim$1-2\%. This difference is largely constant as a function of energy and is therefore an acceptable error. Treating the Crab as a point source allows us to use the built-in pipeline corrections of the Aperture Stop (AS), Ghost Ray (GR) and PSF corrections (the details of these components will be discussed in \S\,\ref{GR_AS}). These corrections are not applicable to extended source responses, and since they are much larger than the error introduced by assuming the Crab is a point source, they are extremely important to include.

We extracted counts from the largest possible source region, typically 200\as, that did not extend beyond the edge of the detector. At this radius about 95\% of all photons are contained in the source extraction region. We used the NuSTARDAS pipeline version v1.3.0 to reduce the data, generate the detector response, RMF, and to apply GR, AS and PSF corrections to the base effective area.


\subsection{Analysis}\label{area_analysis}
Prior to launch, we used ground calibration and modeling to produce a physics-based optics ray-trace model \citep{Westergaard2012} to generate a set of base effective area and PSF files. Figure \ref{beforecrab} shows the ratio of the canonical Crab model folded through the base responses to one of the observations. This comparison shows several features that indicate inaccuracies in the ground calibration:

\begin{itemize}
\item  Below 10\,keV the response turns over exponentially due to uncertainties in the thickness of the Platinum (Pt) electrode on the detector surface.
\item At around 10 keV the Tungsten (W) L-edge causes residuals (See Table \ref{edges}). The exact shape of the W edge depends on the optical constants of the multilayer coatings, which are very sensitive to the crystalline structure and density of the W. Between 10\,keV and 14\,keV the Pt and W L-edge complex, along with the top few coatings of the multilayer stack, create residuals. 
\item Below 20\,keV X-rays reflect off the optics by total external reflection. Above this energy the multilayer coatings become effective and the resonant features, enhancing the reflectivity, are sensitive to the structure of the multilayer that we can only approximate because the coatings are not precisely tabulated for each individual mirror segment of which there are more than 6000.
\item Between 30--40\,keV the effective area deviates gradually from the model areas. The exact reason for this discrepancy is not known.
\item At 69.4\,keV the W K-edge creates a residual for the same reasons as the L-edge. At 78.4\,keV the Pt K-edge is just evident. 
\end{itemize}

In the following sections we model the detected counts in a given instrumental pulse height bin, $C(PI,\theta)$, according to the
equation,
\begin{eqnarray}
C(PI,\theta) &=& \int{{dN(E)\over dE} R(PI,E,\theta) dE}\,.
\end{eqnarray}
Here $dN(E)/dE$ is the model differential photon spectrum of the observed target as a function of incident photon energy, $E$ (flux in units of photons s$^{-1}$ cm$^{-2}$), and $R(PI,E,\theta)$ is the response matrix that captures the photon indicent in a given pulse height bin, $PI$ (units of cm$^2$), at an off-axis angle, $\theta$.  In practice, this integral is approximated as a finite sum by sampling $R(PI,E,\theta)$ on a fine grid.  The off-axis angle, $\theta$, is the angle of incoming X-rays with respect to the optical axis of the mirrors.  As the optical axis moves with respect to the detector position during an observation, the modeled response is sampled on a finite time grid and then summed for a given exposure.

As is typical for X-ray astronomy missions, the response matrix is divided into two components,
\begin{eqnarray} 
R(PI,E,\theta) &=& {\rm RMF}(PI,E) A(E,\theta).
\end{eqnarray}
RMF$(PI,E)$ is known as the redistribution matrix, which contains detector quantum efficiency and resolution effects, and is unitless (a fraction between 0 and 1, where a value of 1 indicates 100\% quantum efficiency). The quantity $A(E,\theta)$ is the effective area, also known as the ancillary response function (ARF), which captures the effective area
of the mirror optics, as well as several other factors unique to \textit{NuSTAR},
\begin{eqnarray} 
A(E,\theta) &=& A_o(E,\theta) V(E,\theta) {\rm detabs}(E) {\rm GR}(E,\theta) \nonumber \\
&&{\rm AS}(E,\theta) C(E,\theta)\,.
\end{eqnarray}
Here $A_0$ is the modeled effective area of the mirror segments estimated from the physics-based ray-tracing simulations, and $V$ is the geometric vignetting function, also based on the ray-tracing simulations. The quantities \textit{detabs}, GR, and AS are the detector dead layer absorption, ghost ray correction, and aperture stop correction, respectively, described in the next sections. $C$ is an empirically derived correction factor, which we will discuss at length in \S\,\ref{empricalcorrection}.

\subsubsection{Detector Absorption Correction}\label{nuabs}
A Pt contact coating and a CdZnTe dead layer on the detector surface cause absorption at low energies. No ground calibration of the absolute efficiency at these energies was performed due to cost and scheduling constraints. Therefore uncertainties  in the low energy detector efficiency remained prior to launch. To investigate this detector absorption, we built an XSPEC absorption model with cross-sections created by Geant4. The adopted photon interaction model is the Livermore low-energy EM model based on the evaluated photon data library, EPDL97 \citep{Cirrone2010}.

Using the Crab complicates the calibration of this absorption layer since the \nh\ column creates a similar effect. At an \nh\ $\sim 3 \times 10^{21}$ cm$^{-2}$ the contribution between 3--4 keV is significant enough that to derive the detector absorption (\textit{detabs}) parameters we had to use 3C\,273, which has a low \nh\ of $1.79 \times 10^{20}$ cm$^{-2}$ \citep{Dickey1990}. At these \nh\ values the quasar 3C\,273 has no measurable absorption above 3\,keV. We fitted only from 3 to 9\,keV, avoiding the W L-edge, and since reflection off the optics at these energies occurs in the regime of total external reflection, where the exact composition of the multilayer stack does not matter, the ground calibration responses were accurate enough for our purpose.

We used XSPEC \citep{Arnaud1996} for the analysis, fitting an absorbed power-law to the data, employing Wilms abundances \citep{Wilms2000} and Verner cross-sections \citep{Verner1996}. Using a power-law index of $\Gamma=1.63$, derived from the cross-calibration campaign to be described in \S \ref{crosscalibration}, we fitted the thickness's of the Pt and CdZnTe for DET0 as shown in table \ref{nuabs} for FPMA and FPMB. Because we only have observations with 3C\,273 on DET0, we used these numbers for all the Crab observations made on DET0 to derive an \nh$=2.224 \times 10^{21}$ cm$^{-2}$ for the Crab. We applied this \nh\ to simultaneous fits of all the Crab observations on each detector respectively for each FPM to derive the other detector dependent absorption parameters. These are summarized in Table \ref{nuabs}.

The measured absorption coefficients are formed into energy-dependent correction files, which are multiplied onto the effective area as an absorption layer in the optical path. These energy correction files, called \textit{detabs}, are stored in the \nustar\ calibration database (CALDB) maintained at NASA HEASARC\footnote{https://heasarc.gsfc.nasa.gov/docs/heasarc/caldb/nustar/}.

\begin{table}[bl]
\centering
\caption{\textit{detabs} parameters}
\begin{tabular}{c|c|c|c}
\hline
Module & Detector & Pt ($\mu$m) & CZT ($\mu$m)\\
\hline
\hline
A & 0 & 0.1143 & 0.2275\\
A & 1 & 0.0993 & 0.3069\\
A & 2 & 0.0712 & 0.3476\\
A & 3 & 0.1081 & 0.2554\\
\hline
B & 0 & 0.1209 & 0.2451\\
B & 1 & 0.0867 & 0.2887\\
B & 2 & 0.1091 & 0.2430\\
B & 3 & 0.0869 & 0.3248\\
\hline
\end{tabular}
\label{nuabs}
\end{table}

\begin{figure}
\begin{center}
\includegraphics[width=0.5\textwidth, bb=100 80 800 450]{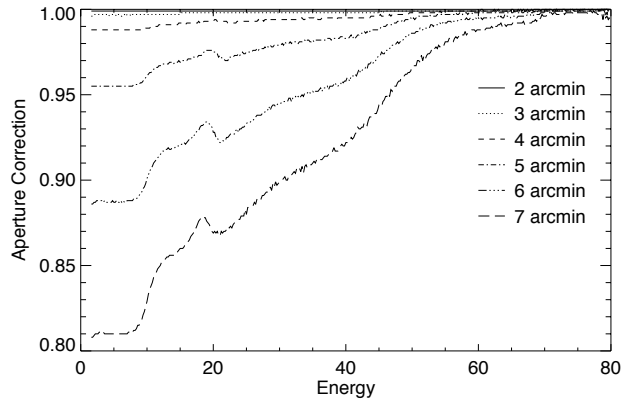}
\end{center}
\caption{Aperture stop correction, AS$(E,\theta)$. The correction is a function of azimuthal location of the source with respect to the optics and off-axis angle. The spectral dependence occurs because the aperture stop preferentially blocks the low energy photons emerging from the outer shells of the optic, which have a larger area.}
\label{APfigure}
\end{figure}

\begin{figure*}[t]
\includegraphics[width=6cm, viewport=50 300 600 750]{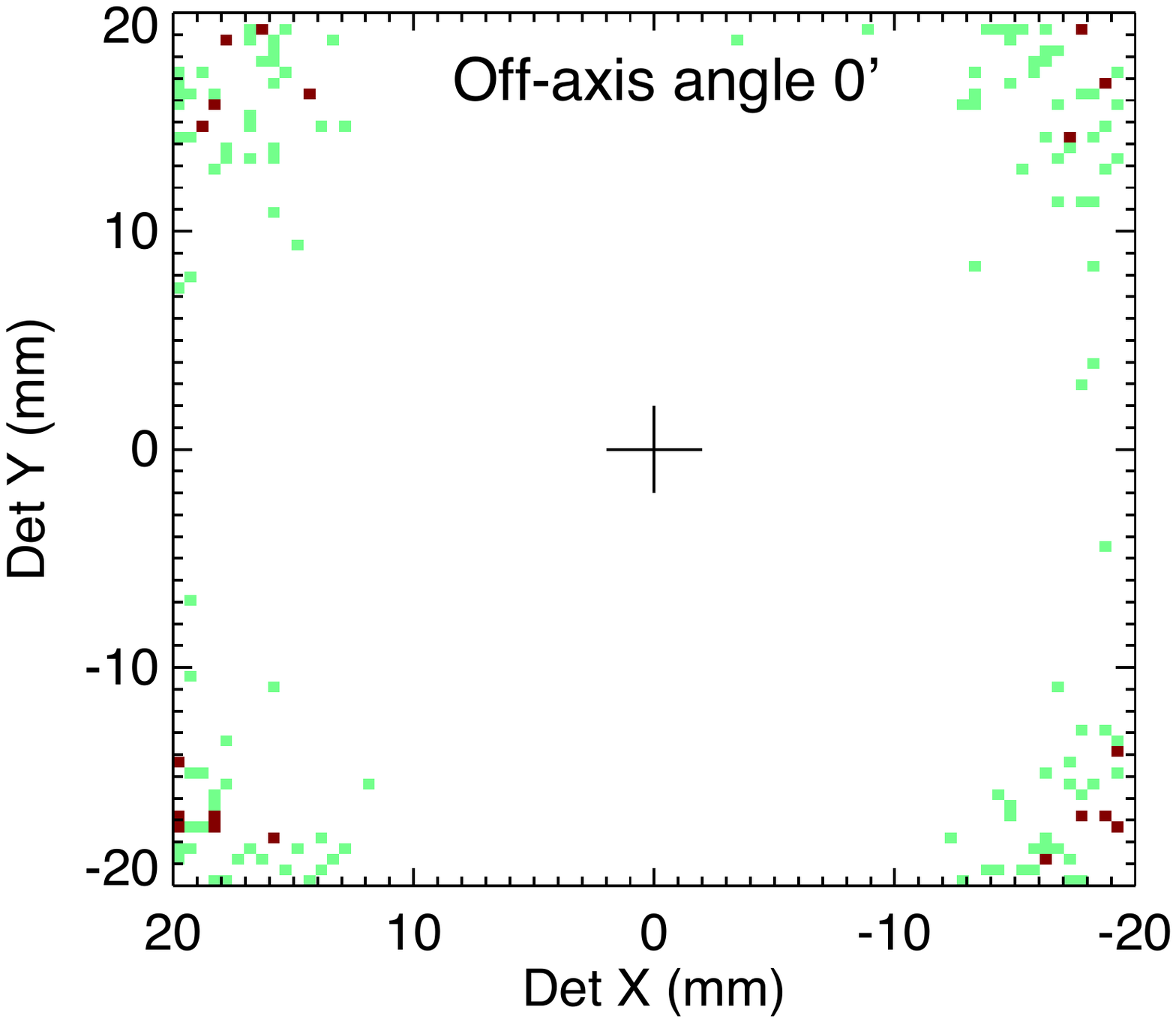}
\includegraphics[width=6cm, viewport=50 300 600 750]{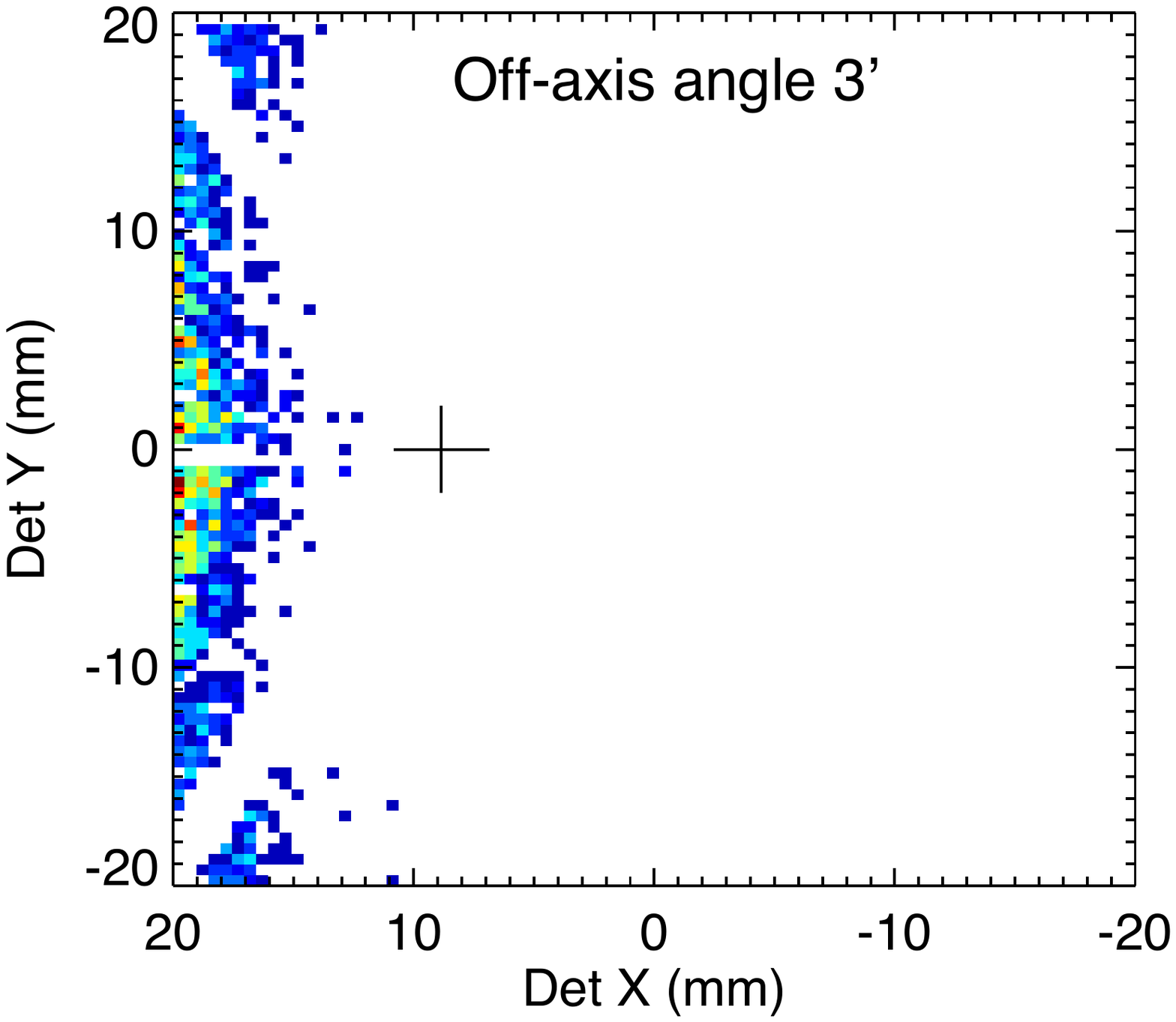}
\includegraphics[width=6cm, viewport=50 300 600 750]{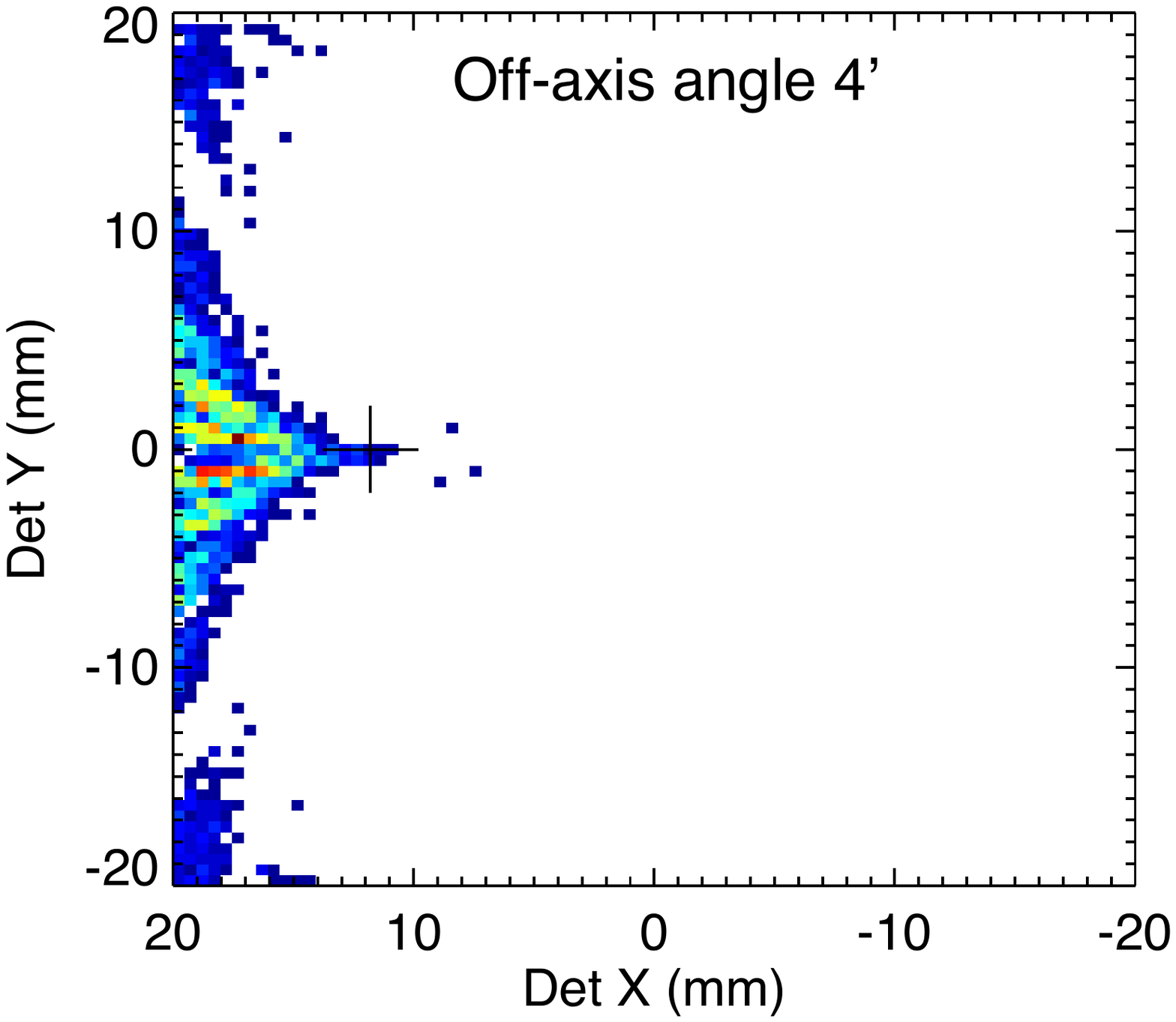}\\
\includegraphics[width=6cm, viewport=50 200 600 750]{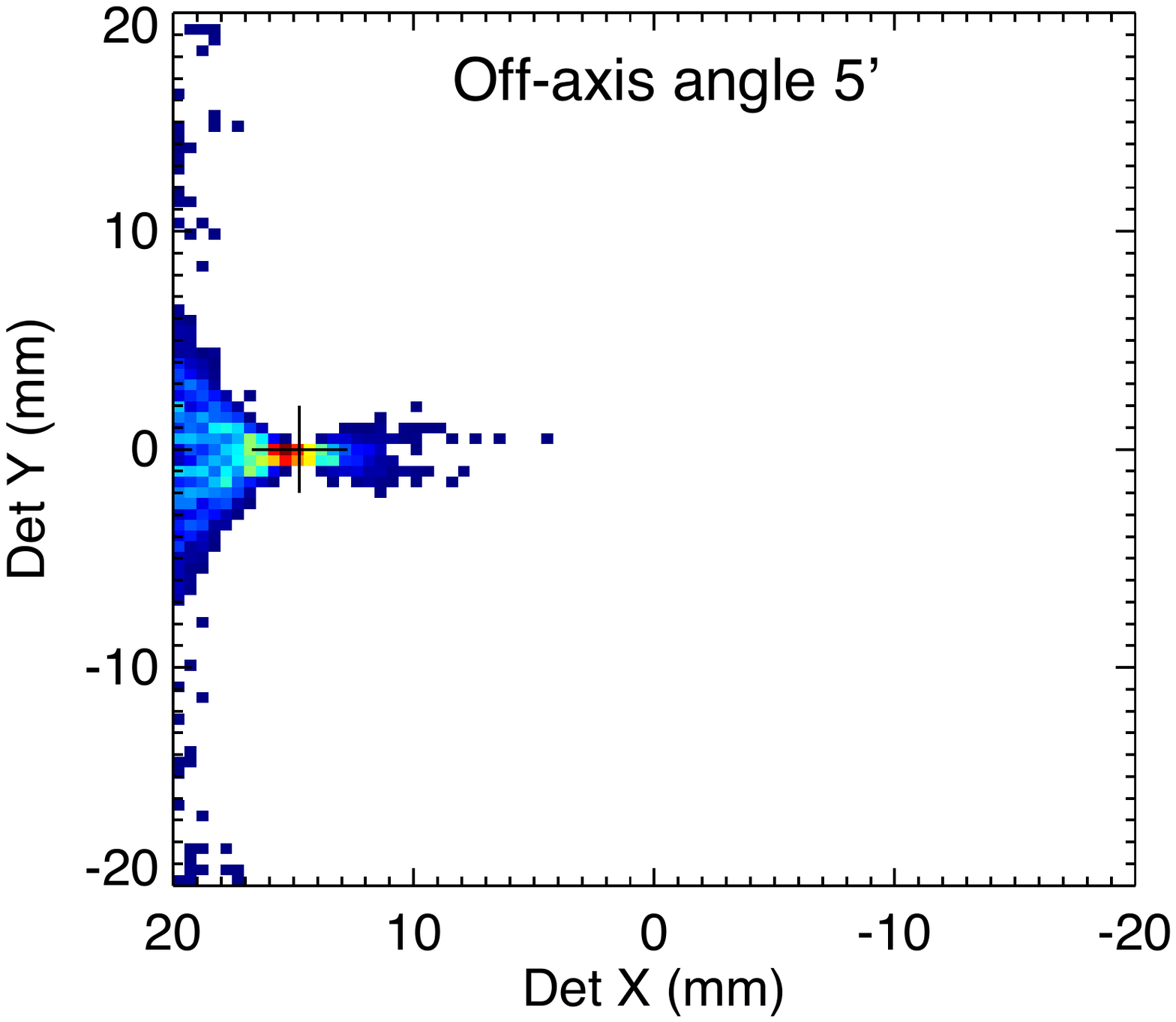}
\includegraphics[width=6cm, viewport=50 200 600 750]{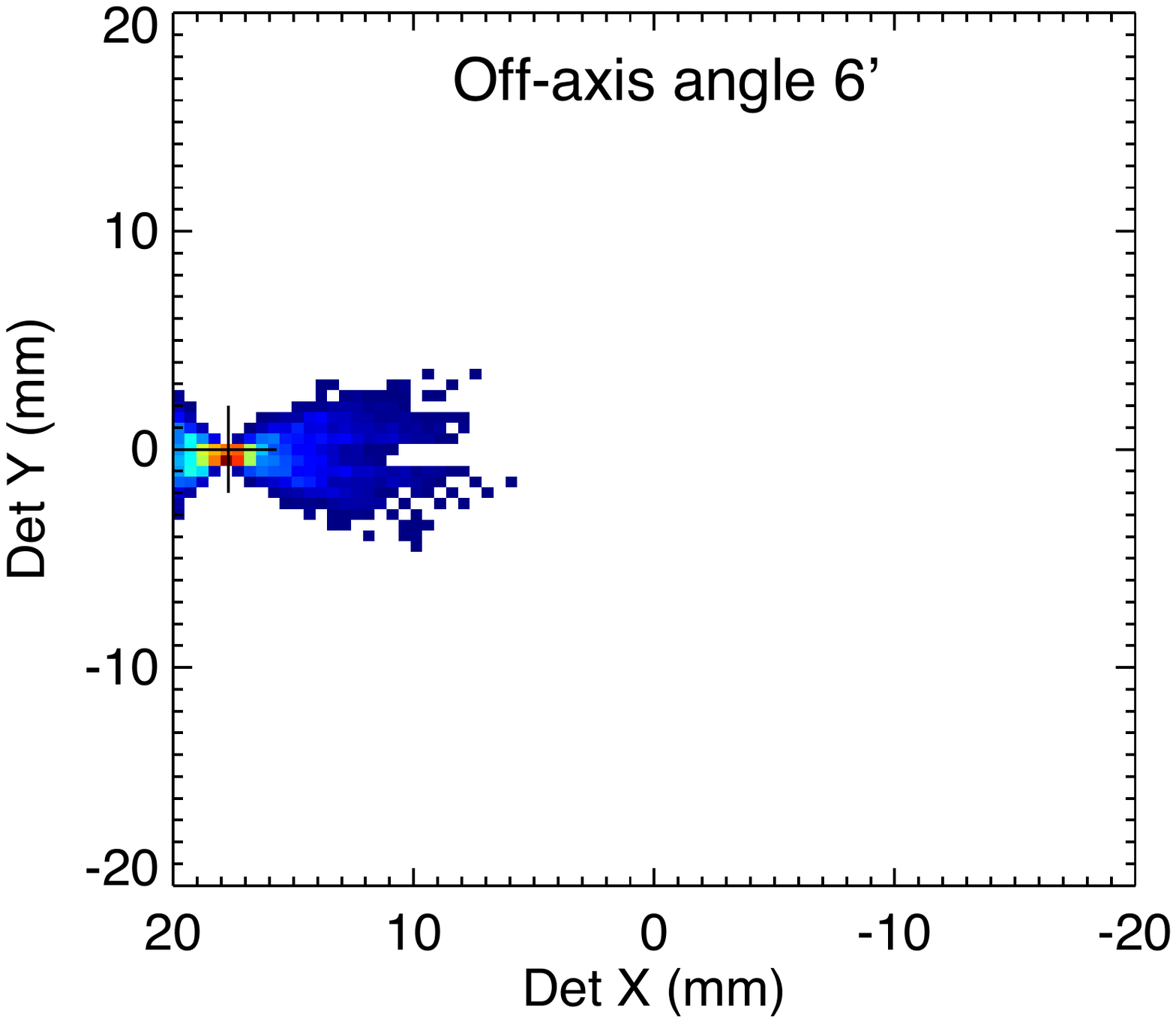}
\includegraphics[width=6cm, viewport=50 200 600 750]{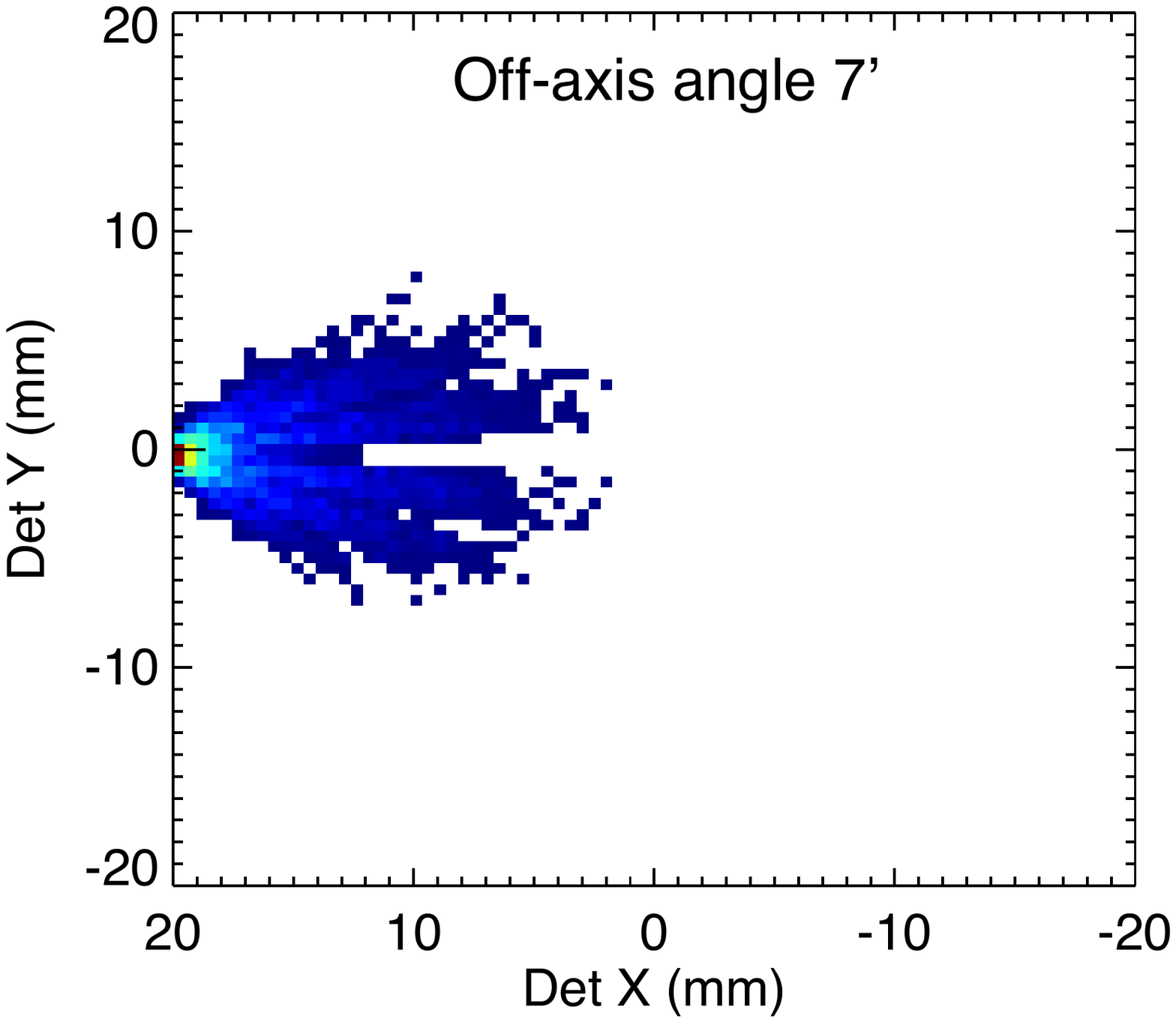}
\caption{Ray-trace Ghost Ray (GR) patterns. Images shows the distribution of GR as a function of off-axis angle. The source location on the detector is initially on-axis and hits the center of the detector marked by the black cross. As the source moves off-axis towards the left the GR population moves towards the right. The peak intensity of the GR initially moves towards the right, but as soon as the source and the GR population meet, the peak intensity of the GR moves together with the source out of the detector's field of view. The petal-like pattern  continues to grow and becomes more diffuse as the source moves further off-axis, until at around 40\am\ reflection of GR is no longer possible. }
\label{GR}
\end{figure*}

\subsubsection{Ghost Rays and Aperture Correction}\label{GR_AS}
The \textit{NuSTAR} observatory has two benches, the Focal Plane Bench (FB) and the Optics Bench (OB) separated by 10.14 meters, which move relative to each other due to motions of the mast that connects them and causes the optical axis to travel across the FPMs. Because the optics act as thin lenses, this track is not equivalent to the path made by the source; the rotational component of a bend in the mast does not project into a movement of the source on the detectors, only the translational part does. In the system there are therefore two paths that must be kept track of as a function of time: the optical axis (OA) path, which is typically on the order of 30\as--60\as\ wide, and the source path, which can be considerably larger and is mainly driven by the pointing stability of the spacecraft bus. The calculation of off-axis angle is carried out as a function of time, and since the resolution of the response files is 10\as, the off-axis aspect histogram is binned at this resolution.

Three aperture stops are attached to each FPM with the top acting as the limiting aperture. The diameter of the aperture opening is 58 mm and the total thickness of the solid part of the aperture is 2 mm, layered into 0.75 mm Al, 0.15 mm Cu and 1.1 mm Sn. The purpose is to block the diffuse background by limiting the field of view (FOV) of the detectors to the open sky. However, this comes at the cost of a reduction in effective area at off-axis angles due to the clipping of the edges of the optical path between detector and optic. Since the aperture stop is fixed with respect to the detector, and the OA is moving independently, the correction is strongly observation dependent. This clipping is a purely geometrical effect and can be calculated by a ray-trace. We have assembled the result as a function of off-axis angle and distance to aperture stop center in a look up table used by the software that creates the ancillary response file for a given observation (\textit{numkarf}). An example of the magnitude of this correction is shown in Figure \ref{APfigure}. The correction is azimuthally dependent since the aperture stop and the OA are not centered, and the correction has a spectral dependence, preferentially cutting away low energy photons since the majority of these come from the outer shells of the optic and thus are more prone to being blocked than light focused by the inner shells. Figure \ref{APfigure} shows that the correction is only important for off-axis angles $>$ 3\am.

\begin{figure}
\begin{center}
\includegraphics[width=0.5\textwidth, bb=100 80 800 450]{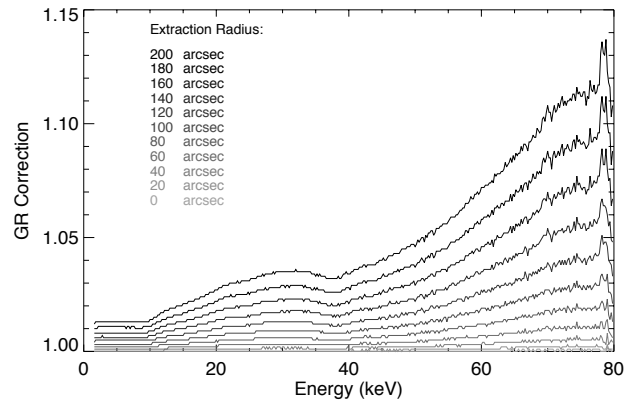}
\end{center}
\caption{GR correction at 4\am. The correction is a function of azimuthal angle of the source with respect to the optics, off-axis angle and extraction region size. It only works with circular extraction regions. The correction increases the effective area.}
\label{GRfigure}
\end{figure}

The \textit{NuSTAR} optics are based on a Wolter-I conical approximation, which is a grazing incidence, double mirror system. A properly focused photon will reflect twice off the optics before exiting. However, it is possible for a photon to reflect only once by either the upper conical section or the lower conical section. This occurs at either very shallow or very steep angle and we call these single-bounce photons Ghost Rays (GR). The pattern of GR is very distinct, as shown in Figure \ref{GR}, and axi-symmetric. Figure \ref{GR} shows the pattern for a series of off-axis angles, starting with the source being on-axis and hitting the center of the detector. In this case the GR are uniformly distributed in a circular halo, which we never see in \textit{NuSTAR} because it is blocked by the aperture stop. As the source moves off-axis and the spot wanders towards the left side of the detector, the GR can be observed entering from the left until around 6\am\ they overlap with the double bounce photons. Since the GR are now inside the typical extraction region, the net effect is an increased effective area. Like the properly reflected photons, these are also subject to being partially blocked by the aperture stop, and so this must also be taken into account. Due to the complexity of the GR pattern, the corrections were only derived for circular regions. Figure \ref{GRfigure} shows an example of the correction, which, like the AS correction, is a function of azimuth, but also depends on source extraction region size and energy. The suppression of the low energies is because of the aperture stop correction to the GR.

The energy dependent correction is applied to the ancillary file in \textit{numkarf}, and because of the complexity of the GR correction files and their dependency on extraction region, this correction is only available for circular and annular regions, and only for point sources. 

\begin{figure}
\centering
\includegraphics[width=0.5\textwidth]{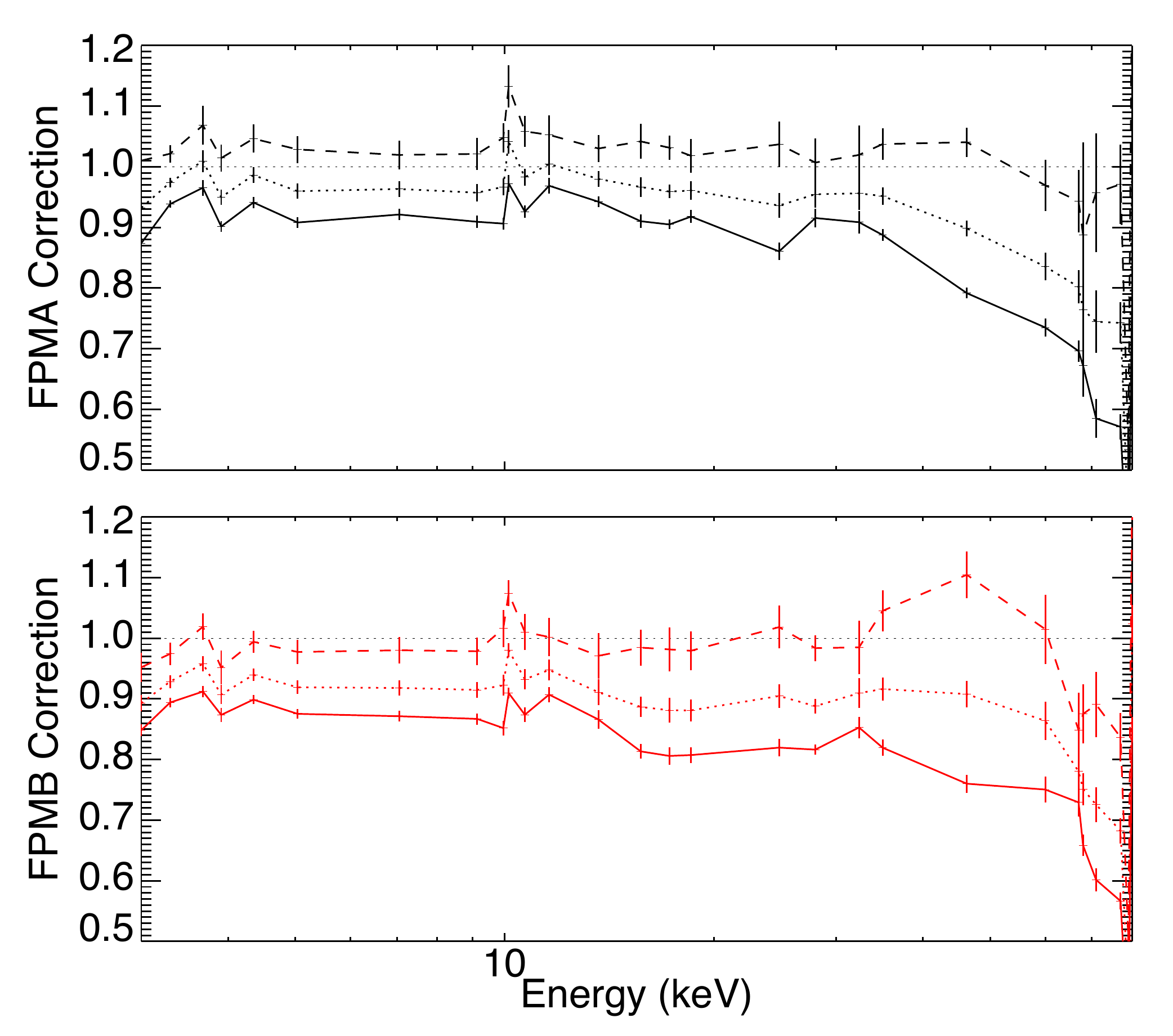}
\caption{Effective area correction factor, $C(E,\theta)$, as a function of off-axis angle and energy for FPMA (top) and FPMB (bottom).  Values are shown for 0\am\ (solid), 3\am\ (dotted) and 7\am (dashed).}
\label{energy_corr}
\end{figure}

\begin{figure}
\centering
\includegraphics[width=0.5\textwidth]{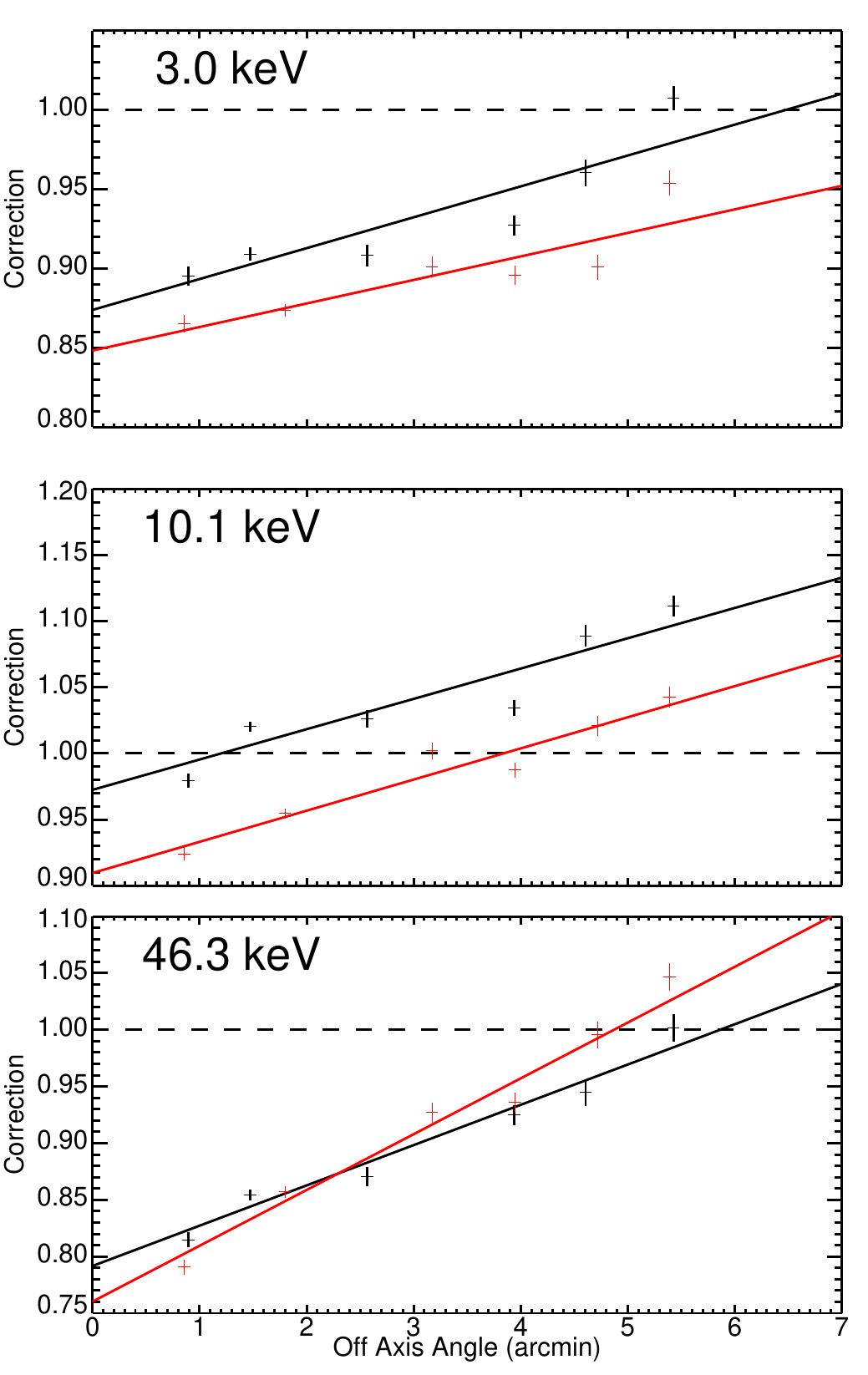}
\caption{Effective area correction factor, $C(E,\theta)$, derived from individual Crab spectra, at representative energies: 3 keV (top), 10.1 keV (middle) and 46.3 keV (bottom).  FPMA is black and FPMB is red.  The thick lines represent the best fit interpolation.  Error bars are standard errors from individual Crab fits, and appear to underestimate the true scatter. Some variability, such as the few percent dip around 3 - -4 arcmin at lower energies, is not captured by the simple linear model.}
\label{offaxis_corr}
\end{figure}

\begin{figure}
\centering
\includegraphics[width=0.47\textwidth]{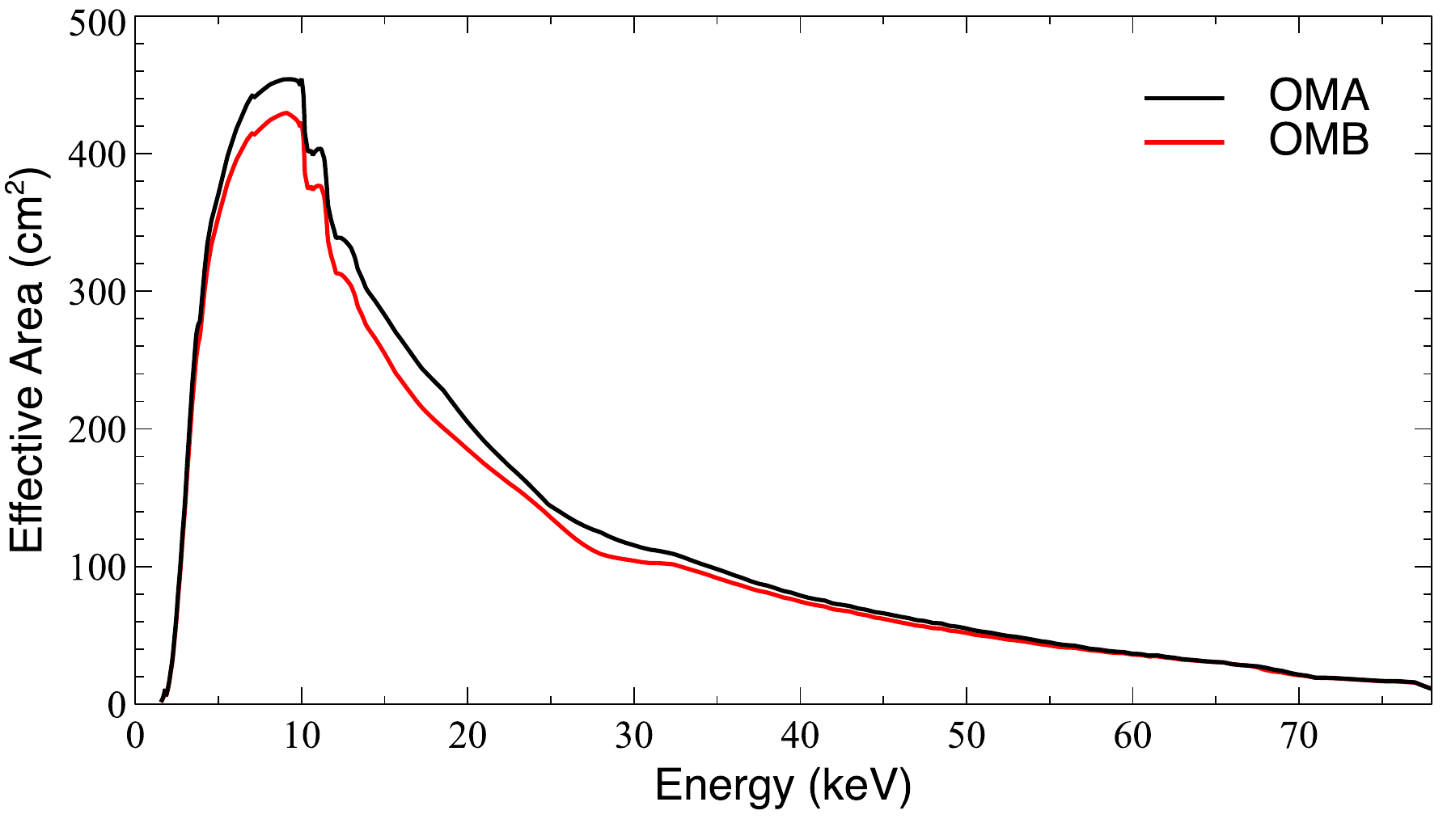}
\includegraphics[width=0.47\textwidth]{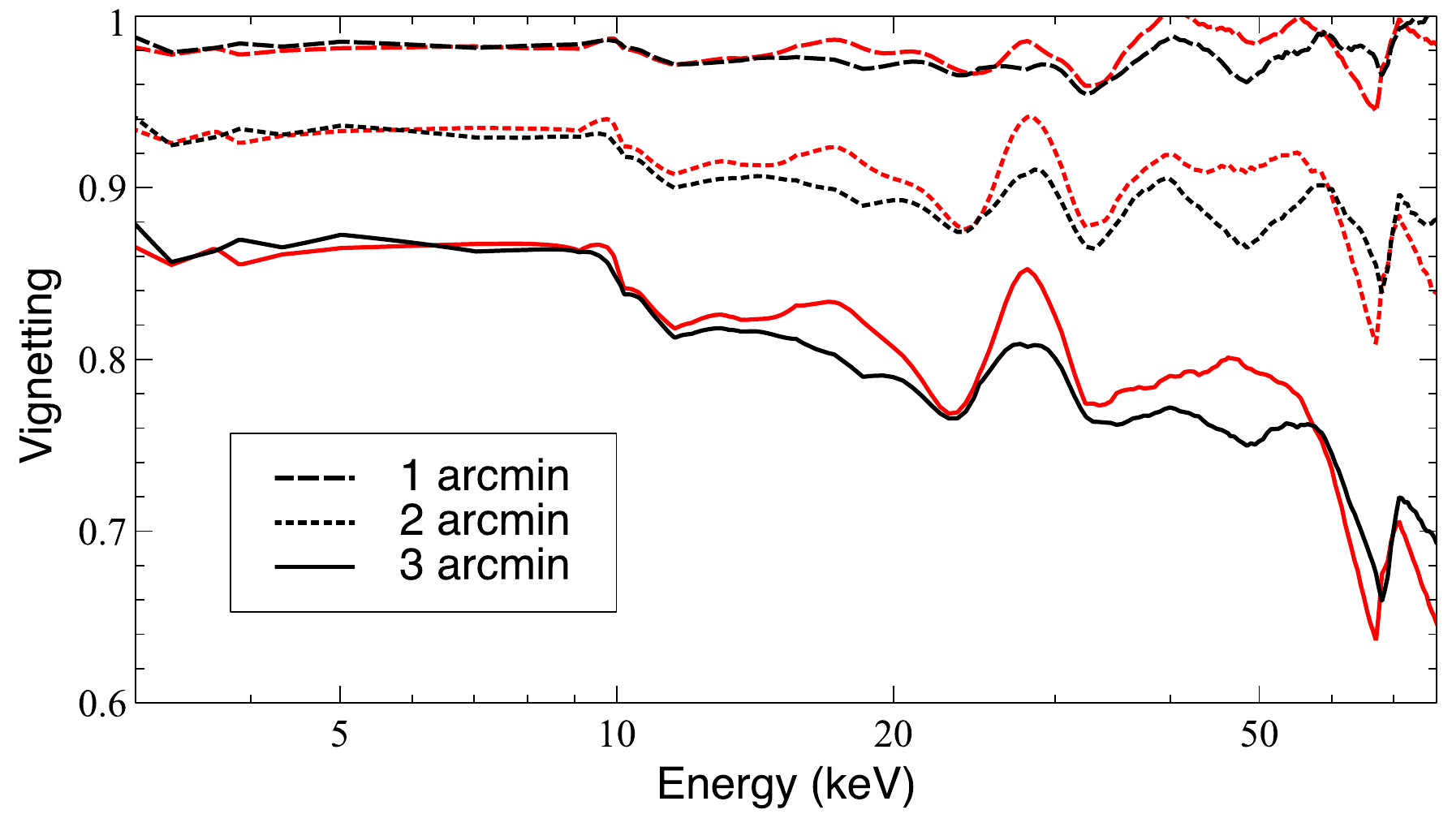}
\caption{Effective Area curves (top) for OMA, and OMB, $A(E,\theta)$, for CALDB file version v006. Vignetting curves (bottom), $V^\prime(E,\theta)$, of OMA and OMB for off-axis angles 1\am, 2\am\ and 3\am\ for CALDB file version v006.}
\label{areanustar}
\end{figure}

\begin{figure}
\begin{center}
\includegraphics[width=0.47\textwidth]{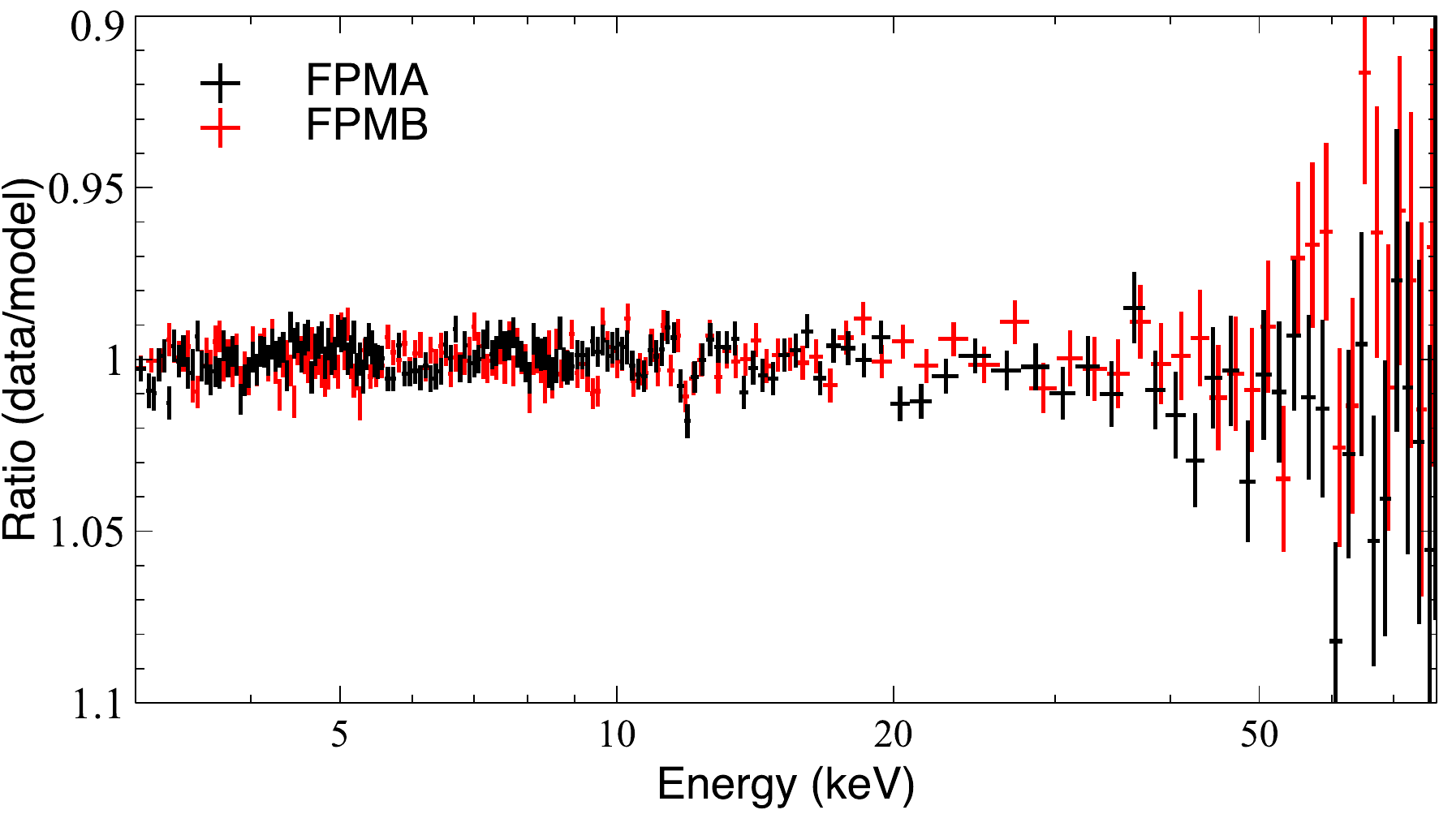}
\end{center}
\caption{Ratio of the base effective area to model ($\Gamma$=2.1, N=10 ph keV$^{-1}$ cm$^{-2}$ s$^{-1}$) for all Crab observations within 3\am.}
\label{aftercrab}
\end{figure}

\subsubsection{Empirical Correction and Vignetting}\label{empricalcorrection}
After adjusting for all the above terms (\textit{detabs}, AS, GR), there still remain systematic residuals in the Crab spectrum which are not accounted for, with typical residuals of a few percent for on axis targets and 10--20\% for off axis targets.  

Although it is desirable to have a completely physics-based response matrix derived from a priori information, this is not usually achievable with limited project resources.  We found that the residual trends as a function of energy, $E$, and off axis angle, $\theta$, are very reproducible, and therefore we elected to adjust the response by an empirical correction factor $C(E,\theta)$, using the canonical Crab spectral model.  $C$ is a fractional quantity, where a value of 1 indicates no correction.  We note that $C$ is a function of energy $E$ (not pulse height), so its effect is equivalent to modifying the effective area term in the ancillary response function (ARF), and its effects must be folded through the response matrix to be seen in the model pulse height spectrum.  As a purely practical matter, $C(E,\theta)$ is stored along with the vignetting function in the \textit{NuSTAR} CALDB (i.e., a new vignetting function, $V^\prime(E,\theta) =V(E,\theta) C(E,\theta)$).

We investigated several functions for $C(E,\theta)$ and decided to use a piece-wise linear interpolation function, in both energy and off-axis angle. This function consists of piece-wise linear segments between fixed control points $E_{ci}$ in energy.  We chose the control points of the spline based on the known features. Figure \ref{energy_corr} marks the locations of the nodes with 27 control points between 3.0 and 80 keV.  The function is completely specified by the value at each of the control points, which are considered the parameters of the correction model. The angular dependence is treated as linear as well, as described further below. We investigated using Chebyshev polynomials or cubic splines as an interpolation function, but encountered difficulty with ringing and other unconstrained behavior in the neighborhood of the control points. The interpolation approximation results shown here worked well for our purposes to within available statistics of the data.  An improved model may be found in the future, but the present correction achieves residuals at the few percent level over almost the entire \textit{NuSTAR} energy range.  

We derived the correction spectrum in two stages.  First, we fit individual Crab spectra to derive the energy dependence for several off-axis angles.  Second, we used linear interpolation to model the angular dependence to derive a single global correction model.

For the first stage of analysis, we combined the observations in 1\am\ bins to improve statistics, resulting in seven spectra between 0\am\ (on-axis) and 7\am\ off-axis.  We fit the parameters of the correction model while holding the canonical spectral model fixed, using standard $\chi^2$ fitting, and requiring at least 200 counts per bin.  The result is a linear model for each of the seven spectra.  We note that for the highest control points at 77 and 80 keV, there were not enough counts to reliably estimate the parameter values at both points independently, so the value of the response at the 80 keV control point was fixed manually to be 90\% of the value at 77 keV.  In the final result, we claim calibrated flux levels only between 3--78 keV.

The corrections show an angular dependence nearly linear with off-axis angle (Figure \ref{offaxis_corr}).  This indicates that the physics-based ray-tracing captured most of the angular dependence of the reflectivity of the mirrors, except for a residual trend.  We therefore treat the angular dependence of the correction as linear between on-axis and 7\am\ off-axis.  For off-axis angles above 7\am\ (the outer parts of the focal plane) we lack the Crab data to constrain the model, so we cannot verify if the linear trend continues.  To be conservative, we do not extrapolate the linear trend beyond 7\am, but simply hold the function fixed at the 7\am\ value.

In a second stage of analysis, we fitted the individual correction values at each control point independently with a linear function, using the standard errors from the individual fits, to derive one global correction function for each FPM.  We stress that this global correction is an ensemble-derived model using all of the Crab observations; no single Crab spectrum was used to fix the final global model parameters.  Thus, the final global correction model consists of 28 correction parameters on-axis and 28 parameters 7\am\ off-axis, for a total number of model parameters of 56 for each FPM.  These are derived from $7\times 26$ (=182) individual correction spectra, which were in turn derived from $7\times 940$ ($\simeq$ 6600) individual Crab count bins for each FPM.  Figure \ref{energy_corr} shows the correction function $C(E,\theta)$ as a function of energy $E$ and off-axis angle $\theta$ for both FPMs. The systematic nature of the adjustments and some of the features mentioned in \S\,\ref{area_analysis} are evident.

As noted above, the correction factor $C(E,\theta)$ and geometric vignetting function derived from ray tracing $V(E,\theta)$ were combined into a single effective vignetting function $V^\prime(E,\theta)$, and stored in the \textit{NuSTAR} CALDB.  This effective vignetting function is shown in Figure \ref{areanustar} (bottom), along with the resulting total effective area for each module (top).

We reprocessed the Crab data with the adjusted calibration files to verify that the correction was successful.  Figure \ref{aftercrab} shows the residual spectrum for all Crab data within 3\am\ of on-axis for both modules. The residuals are typically better than $\pm$2\% up to $\sim$40 keV.  Between 40--80 keV, residuals are dominated by counting statistics of the Crab data, but are typically 5--10\%, so the systematic errors are less than that.

The final correction achieves few percent residuals over the entire \textit{NuSTAR} energy range of 3--78 keV, and is applicable for off-axis angles of 0\am--7\am. We do not advise to fit spectra above 78 keV.

\begin{figure}
\begin{center}
\includegraphics[width=0.47\textwidth]{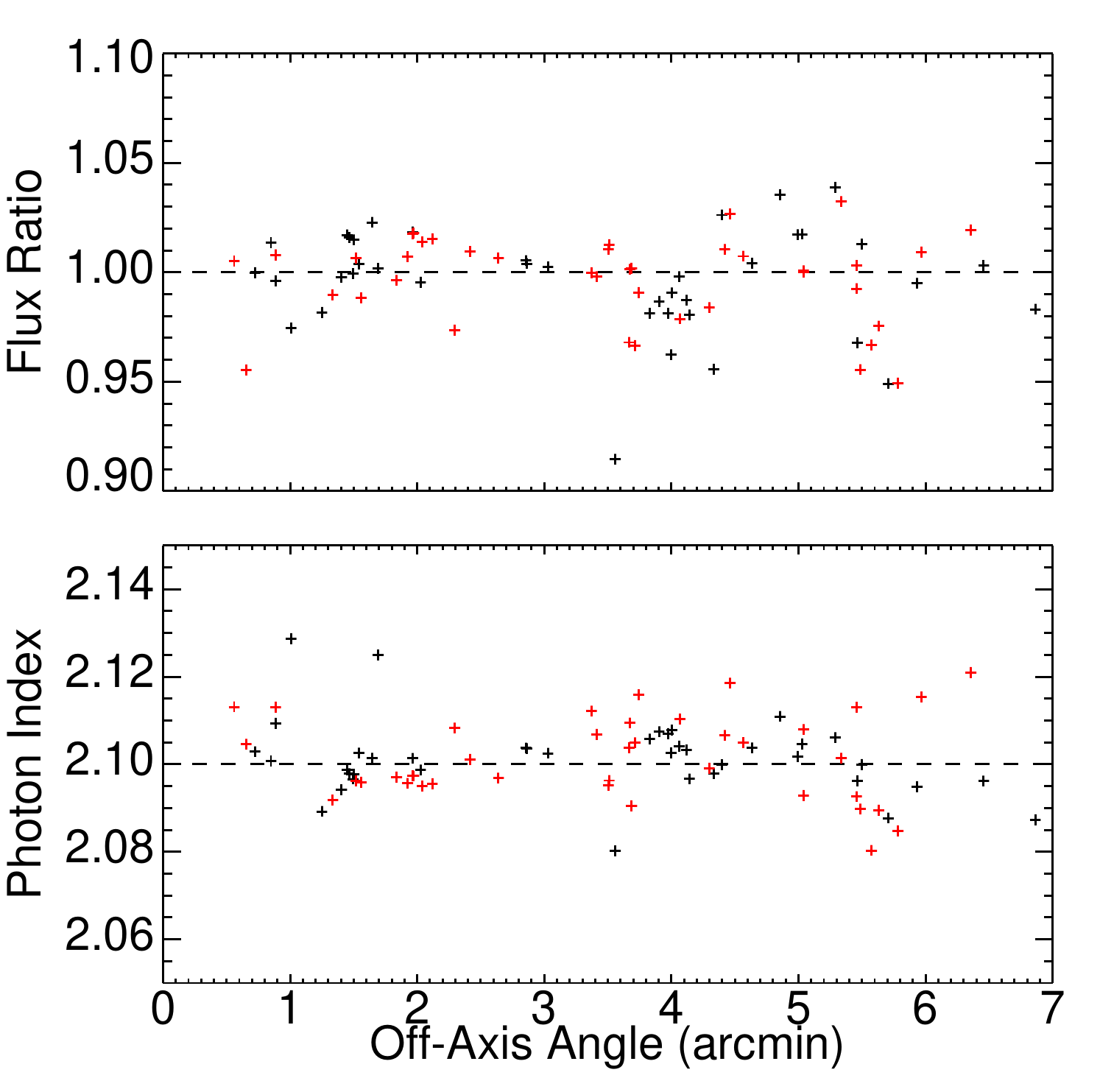}
\end{center}
\caption{Residual spectral parameter values for relative flux (top)
  and power law photon index (bottom) as a function of off-axis angle,
  $\theta$.  FPMA is black and FPMB is red.  The canonical values are
  shown as dashed lines.}
\label{flux_vs_angle}
\end{figure}

\begin{table}
\centering
\caption{Measurement Repeatability Statistics}
\begin{tabular}{lcc}
\hline
Subset                   & Mean & Std. Dev.\\
\hline
\hline
\multicolumn{3}{c}{Relative Flux}\\
\hline
All                       & $-$0.4\% &  2.6\%\\
On-Axis  ($\theta<$3')    & $-$0.9\% &  3.0\%\\
Off-Axis ($\theta>$3')    & $-$0.9\% &  3.0\%\\
DET0                      & $-$0.3\% &  2.5\%\\
DET1                      & $-$0.7\% &  2.5\%\\
DET2                      & $+$0.8\% &  1.4\%\\
DET3                      & $-$0.0\% &  3.4\%\\
DET0+3 Straddle           & $-$1.7\% &  4.0\%\\
FPMA$-$FPMB Diff.         & $-$0.0\% &  2.6\%\\
\hline
\multicolumn{3}{c}{Power-Law Photon Index, $\Gamma$}\\
\hline
All                       & ---    &  0.009\\
FPMA$-$FPMB Diff.         & ---    &  0.012\\
\hline
\multicolumn{3}{c}{Absorption, \nh  (10$^{22}\,{\rm cm}^{-2}$)}\\
\hline
All                       & --     &  0.150\\
\hline
\end{tabular}
\label{crab_spect_stats}
\end{table}

\subsection{Measurement Repeatability}\label{repeatability}
Having calculated the correction function, as a consistency check we reprocessed all individual Crab data sets to determine if the specified Crab spectral parameters were recovered.  Although the Crab spectral shape and normalization were enforced at the ensemble level during the correction process, we did not required that individual spectra conform, nor did we require that FPMA and FPMB produce the same results.  Any deviations at the individual observation level would indicate unmodeled systematic errors, as well as errors in the assumption of a constant Crab.  To estimate this, we computed sample biases and standard deviations for various subsets of observations. This is a measure of how repeatable \textit{NuSTAR} observations are.

We fit each of the spectra with the canonical spectral shape, but allowed the parameters to vary.  The parameters were the relative power-law flux normalization and the power law photon index, $\Gamma$, while the neutral absorption parameter, \nh, was fixed.  We also performed a separate fit where in addition \nh\ was allowed to vary. Results from FPMA and FPMB were recorded separately.

We recorded the individual parameters and computed statistics of various subsets of observations.  The results are shown in Table \ref{crab_spect_stats}.  In addition to the sample variation among ``All'' observations, we also examined subsets where the target was near on-axis and far off-axis, and on individual detectors (DET0--3).  These designations are not mutually exclusive. In a few cases the Crab straddled two detectors, and these are shown separately.  We also computed the sample variations of the differences between FPMA and FPMB. For relative flux, we computed the sample bias as well as the standard deviation; for photon index and absorption, we only list the sample standard deviation. The results are also shown graphically in Figure \ref{flux_vs_angle} as a function of off-axis angle.

The results demonstrate that average relative flux biases are small, typically $<$1\%.  Scatter in the relative fluxes is typically 2--3\% (1$\sigma$).  There are no large flux offsets between FPMA and FPMB during the same observation; in the worst case the difference was 3.8\% in this sample.  There are no strong biases as a function of off-axis angle or detector number. The power law index was stable to about $\pm$0.01 (1$\sigma$); this can be taken as an approximate estimate of repeatibility errors in spectral slope. For neutral absorption, \textit{NuSTAR} is not particularly sensitive to the Crab's hydrogen column; the sample standard deviation is comparable to the canonical value itself.

The fluctuations in flux and slope are understood, and are consistent with the estimated error in optical axis location. The knowledge of the optical axis location is accurate to $\sim 30$\as, and because the optical axis is constantly moving due to the mast motions, there will be situations where the agreement between effective areas are better than in others. The flux level is also affected by the understanding of the PSF (to be discussed in the next section), and the fidelity of the instrument map; gaps, dead pixels etc. The gap width between detectors was measured on the ground by scanning an X-ray beam across the gaps, but due to systematic errors getting the exact physical gap width is challenging, making it difficult to correct the flux exactly in the cases where the core of the PSF falls, or crosses the gap. 

Observed differences between the two modules of up to 5\% are not uncommon and are simply a result of the optical axis configuration during the time of observation. In extension, observed flux differences less than 5\% between different observations of the same target, should not be interpreted as intrinsic to the source.

\begin{table}[b]
\centering
\caption{\nustar PSF Observations log}
\begin{tabular}{lccc}
\hline
Target & Obs ID & Exposure & Off-axis angle\footnote{Average value for the distribution} \\
& & (ks) & (\am) \\
\hline 
\hline
Cyg~X-1 & 10002003001 & 9 & 0.5 \\
GRS~1915+105 & 10002004001 & 15  & 1  \\
Vela X-1 & 10002007001 & 11 & 1 \\
Vela X-1 & 30002007002 & 7 & 1 \\
Vela X-1 & 30002007003 & 24 & 2 \\
GS0834-430 & 10002018001 & 31 & 1 \\
Her~X-1 & 30002006002 & 28 & 2 \\
Her~X-1 & 30002006005 & 22 & 1 \\
Her~X-1 & 30002006007 & 27 & 1 \\
Cyg~X-1\footnote{for off-axis PSF} & 00001007001 & 2.4  & 3 \\
Cyg~X-1$^{\rm b}$ & 00001008001 & 4.3  & 3 \\
\hline
\end{tabular}
\label{ta:psfobsid}
\end{table} 

\begin{table}
\centering
\caption{Measured half power diameter for the on-axis observations in Table~\ref{ta:psfobsid}}
\scriptsize{
\begin{tabular}{ccccc|cccc}
\hline
Energy & \multicolumn{4}{c|}{FPMA} & \multicolumn{4}{c}{FPMB} \\ \hline
 & HDP & $\sigma_{\rm HPD}$ & min. & max. & HPD & $\sigma_{\rm HPD}$ & min. & max.  \\
(keV) & ($''$) & ($''$) & ($''$) & ($''$)& ($''$) & ($''$) & ($''$) & ($''$) \\
\hline
\hline
3--4.5   & 70.3 & 2.4 & 66.7 & 75.5 & 65.6 & 2.4 & 62.9 & 69.6  \\
4.5--6   & 67.1 & 1.0 & 64.7 & 67.7 & 62.6 & 1.2 & 60.9 & 64.7  \\
6--8     & 64.7 & 1.0 & 62.8 & 65.7 & 60.7 & 1.4 & 58.8 & 63.7  \\
8--12    & 63.5 & 1.1 & 61.8 & 64.7 & 59.5 & 1.5 & 57.9 & 62.8  \\
12--20   & 63.4 & 1.1 & 61.8 & 64.7 & 60.3 & 1.2 & 58.8 & 62.8  \\
20--79   & 63.4 & 1.0 & 61.8 & 64.7 & 62.4 & 1.5 & 60.8 & 65.7  \\
\hline
\end{tabular}}
\label{ta:hpdvals}
\end{table}

\section{Point Spread Function Calibration}\label{psf}
\subsection{Near On-Axis PSF Shape}
Calculating the correct effective area requires a good understanding of the PSF. Prior to launch a ray-trace model of the PSF was developed and tested against limited ground calibration scatter data from individual flight mirrors \citep{Westergaard2012}. This original set of calibration files formed the basis of pre-launch PSFs determined every 0.5\am\ out to an off-axis angle of $\theta=8.5'$. 

To conduct the in-orbit calibration of the PSF, we observed the strong point sources listed in Table~\ref{ta:psfobsid}. The main point of these observations was to establish the corrections required to the ground based estimates. To this end, we extracted events from the point source observations in several energy ranges (3--4.5, 4.5--6, 6--8, 8--12, 12--20, and 20--78\,keV) and produced radial profiles centered at the peak of the intensity map. Table~\ref{ta:hpdvals} shows the half power diameters (HPD) for the nine on-axis observations.

We compared these radial profiles with the ray-trace model. Since the off-axis angle changes during each observation due to the mast motions, we weighted the ray-traced PSF with the aspect solution before comparing. We fitted the observed radial profiles to the aspect-weighted ray-trace PSF and found that: (i) the central cores of observed radial profiles are slightly broader than that of the model PSF, (ii) there is an unaccounted for wing at $R>100''$ in the radial profiles. The broadening of the central core does not affect the FWHM and is an effect of incomplete astrometric aspect reconstruction. The origin of the wider wing is unknown.

To take care of the central broadening we convolved the modeled PSF with a Gaussian function. Since the origin of the wing is unknown, we decided to empirically model it by introducing a component on top of the ray-trace PSF. Because background subtraction is not feasible during the fitting, we had to include a background component as well, which is well known and can be modeled accurately \citep{Wik2014}. The backgrounds were in general small compared to the sources, dominating only above $R>500''$. 

The 2-D model PSF can be expressed by the following formula:
$$\Psi(x,y)=C_0\Psi_{\rm RT}(x,y)\otimes G(x,y,\sigma)$$
$$ + C_1 e^{-a\sqrt{x^2 + y^2}} +C_2B(x,y)$$
where $\Psi_{\rm RT}$ is the ray-trace PSF, $G(x,y,\sigma)$ is the Gaussian function, $e^{a\sqrt{x^2 + y^2}}$ is the wing component, $B(x,y)$ is the background model, and $C_{0,1,2}$ are normalization constants.

\begin{figure}
\centering
\includegraphics[width=0.5\textwidth]{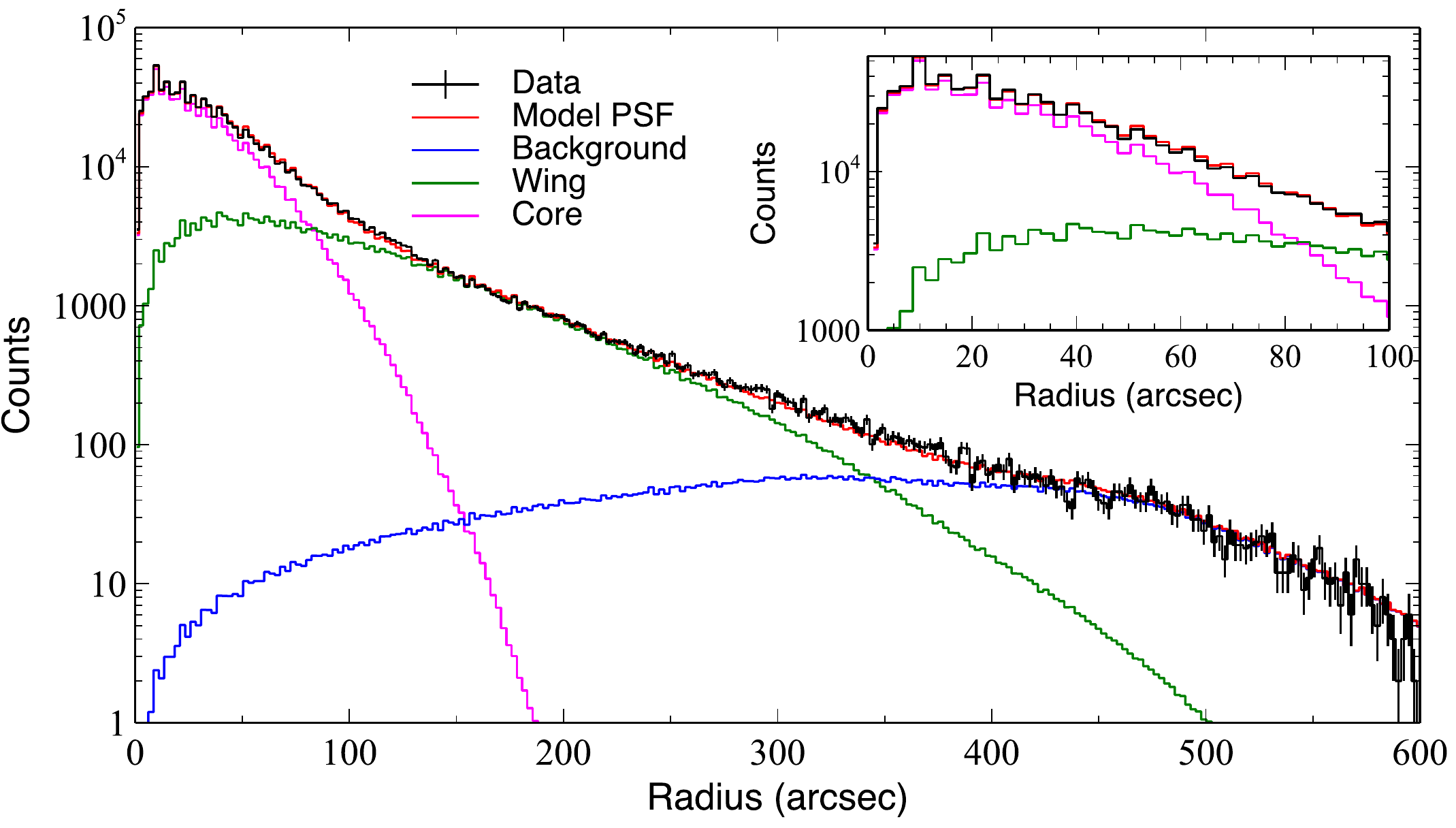}
\caption{Radial profile of Cyg\,X-1 in the 5--8\,keV band. Fit model components are also shown.
Background components were fit separately but only the combined model is shown in red.}
\label{fig:psffit}
\end{figure}

To fit the radial PSF profiles, we first convolved the ray-trace PSFs with the Gaussian function and weighted them using the aspect solution of each observation to produce an aspect-weighted PSF. The wing model was added to construct an effective PSF, and we multiplied it with the exposure map for the observation. Since the point source exposure map is almost flat, the overall shape of the PSF does not change much. However, multiplying with the exposure map is important for some localized effects such as detector gaps. 

The fitting parameters are the amplitude of the ray-trace PSF ($C_{\rm 0}$), width of the Gaussian convolution core ($\sigma$), amplitude and decay constant for the exponential wing ($C_{\rm 1}$ and $a$), and amplitude of the background ($C_{\rm 2}$). Since fitting the observations simultaneously was difficult given different aspect histories (i.e., different effective PSFs) and backgrounds, we calculated $\chi^2$ for various values of fitting parameters for each observation, and found a set of parameters that minimize the combined $\chi^2$. An example the observed radial profile for a Cyg\,X-1 observation (obs. ID: 10002003001) and the best-fit PSF model is shown in Figure~\ref{fig:psffit}.

\begin{figure}
\centering
\includegraphics[viewport=80 50 830 600,width=0.57\textwidth]{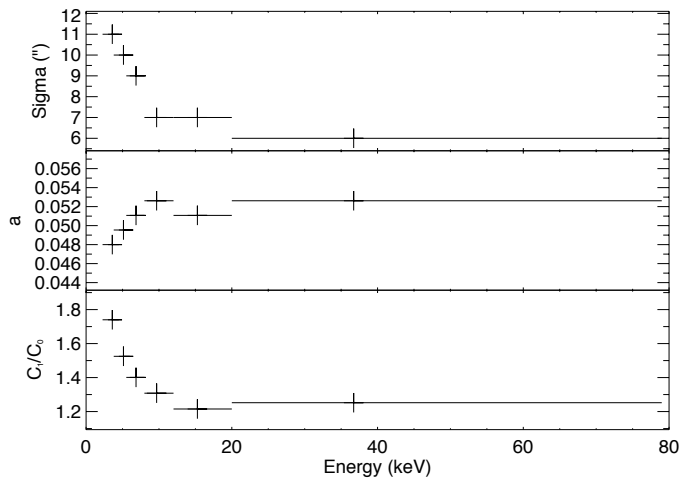}
\caption{Measured PSF fitting parameters as a function of energy.}\label{fig:pars}
\end{figure}

\begin{figure}
\centering
\includegraphics[viewport=30 50 830 600,width=0.57\textwidth]{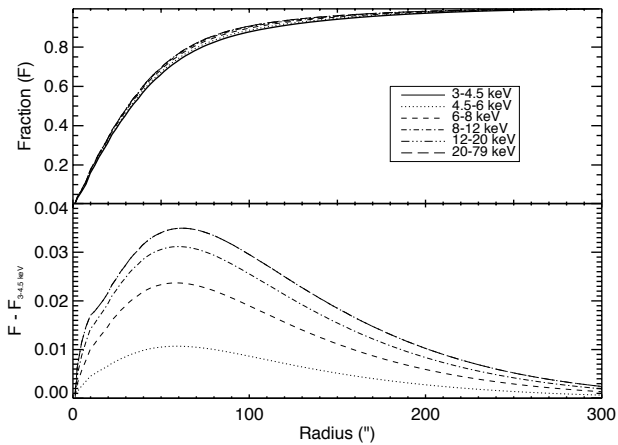}
\caption{Enclosed counts fraction in different energy bands (top) and the differences from
that in the 3--4.5\,keV band (bottom).}\label{fig:encl}
\end{figure}

From the analysis, we found that the PSF sharpens - the half power diameter decreases - with energy. As a function of increasing energy we find a smaller core-convolution width and narrower wing, as well as a relatively larger amplitude for the core component (see Figure \ref{fig:pars}). The sharpening saturates at $\sim$10\,keV. The differences in the enclosed count fractions at different energies are modest, with a maximum of $\sim$3\% between the lowest and the highest energy bands for a $\sim$1\am\ extraction radius region as shown in Figure \ref{fig:encl}. However, since this is significant enough for small extraction regions, we include the energy dependent evolution into the ARF generation, linearly interpolating between energies. The additional broadening of the wings of the PSF at energies below 10\,keV is likely explained by scattering from surface contaminants.

\begin{figure}
\centering
\includegraphics[width=0.5\textwidth]{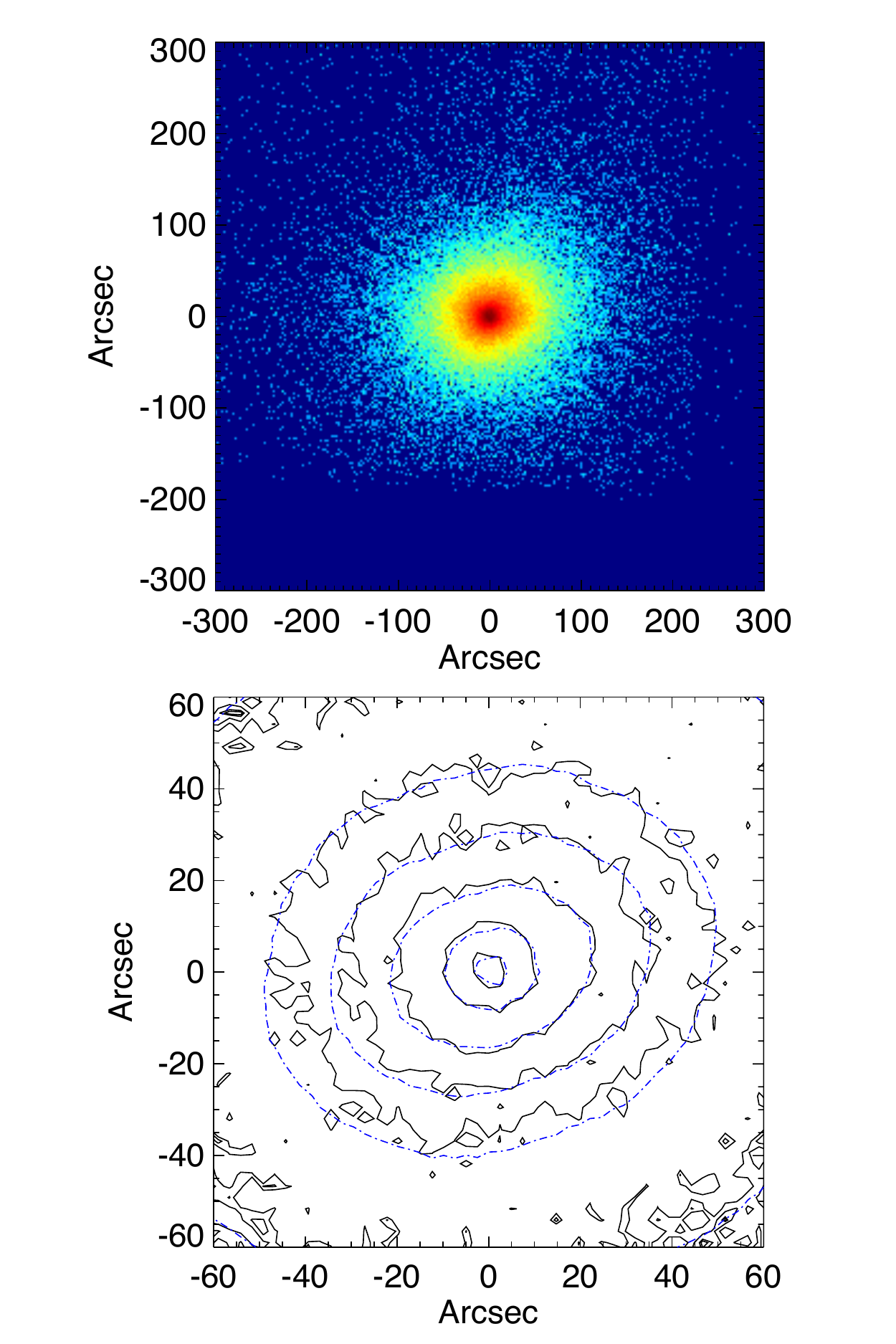}
\caption{Upper: Source image. Lower: Count contours (solid) and PSF contours (dashed) for a $\sim$3\am\ off-axis observation of the Cyg~X-1 in a $2'\times2'$ field.}
\label{fig:off}
\end{figure}

\begin{figure}
\centering
\includegraphics[viewport=30 50 830 600,width=0.57\textwidth]{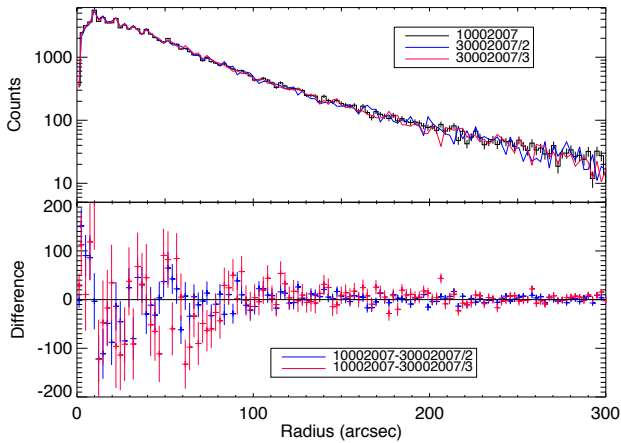}
\caption{Observed 3--4.5 keV PSFs (top) for two Vela~X-1 separated by $\sim$300 days and the difference
with 1$\sigma$ statistical errors (bottom). The PSF at T=0 is shown in black, and PSFs measured at T=300 days
are shown in blue and red.}
\label{fig:PSFtime}
\vspace{3mm}
\end{figure}

\begin{table*}
\centering
\caption{Gain-Corrections (2012-vs-2015)}
\begin{tabular}{lrrrr}
\hline
Detector Number & A Slope & A Offset (eV) & B Slope & B Offset (eV) \\
\hline
0 & 0.9944 &  -60 & 0.9964 & -59 \\ 
1 & 0.9928 & -66 & 0.9939 & -56 \\
2 & 0.9999 & -34 & 0.9937 & -34 \\
3 & 0.9977 &  -55 & 0.9963 & -7 \\
\hline
\end{tabular}
\label{gainfits}
\end{table*}

\begin{table*}
\centering
\caption{Cross-calibration Campaign Observations}
\begin{tabular}{l||c|c||c|c}
\hline
Observatory & OBSID & Exposure & OBSID & Exposure \\
            &       & (ksec)   &       & (ksec) \\
\hline
& \multicolumn{2}{c}{PKS2155-304} & \multicolumn{2}{c}{3C\,273} \\
\hline
\textit{NuSTAR} & 60002022002 & 45 & 10002020001 & 244\\
\textit{Chandra} & 15475 & 30 & 14455 & 30\\
\textit{XMM-Newton} & 0411782101 & 76 & 0414191001 & 38.9\\
\textit{Swift} & 00030795108 & 17.7 & 00050900019	& 13\\
\textit{Swift} & -- & -- & 00050900020	& 6.9 \\
\textit{Suzaku} & 108010010 & 53.3 & 107013010 & 39.8 \\
\hline
\end{tabular}
\label{crossobsid}
\end{table*}

\subsubsection{Off-Axis PSF}\label{offpsf}
The FWHM of the \textit{NuSTAR} PSF does not change with off-axis angle, but the shape of the PSF gradually distorts azimuthally, and the geometrical shadowing of the shells causes the PSF to appear elongated, as shown in Figure \ref{fig:off}. Probing the PSF at high off-axis angles is complicated by the lower count-rate, and necessitates a 2-D approach where we match the contours of the model PSF to the actual PSF.

We followed the same procedure (e.g., convolution, aspect-weighting, etc) as we did for the on-axis studies to produce the model PSF for the off-axis observations, $\theta>3'$ (listed in Table \ref{ta:psfobsid}). The ray-trace PSF includes the effect of geometrical shadowing (vignetting of photons), and successfully reproduces the observed PSF at the core. However, the wing component, which dominates at R$>$100\as, did not match with the observed contours at large radii. We parametrized the distortion of the PSF wing with an ellipticity factor and a shadow function, where the form of the shadow function was derived using the geometry of the optics:
$$F_{\rm shadow}(x,y, \theta)=1-A(\theta){\rm cos^{2}}({\rm tan^{-1}}(B(\theta)y/x)),$$
$$F_{\rm wing}(x,y, \theta)=e^{-a\sqrt{x^2 + B(\theta)y^2}},$$
where $A(\theta)$ is the shadow factor, $B(\theta)$ is the ellipticity factor, and $\theta$ is the off-axis angle in arcminutes. The shadow function was multiplied onto the wing component, which was added to the ray-trace PSF. The parameters were determined using the off-axis point source observations to match the 2-D contours at large radii, and we found that $A(\theta)=0.025\theta$ and $B(\theta)=1-0.025\theta$ made the contours match at large radii (see Figure \ref{fig:off}).

With these correction factors we generated new energy-dependent 2-D PSF files for the \textit{NuSTAR} CALDB at  3--4.5, 4.5--6, 6--8, 8--12, 12--20, and 20--78\,keV. 

\subsubsection{PSF Stability over Time}\label{sec:psftime}
Changes in PSF over time can occur from out-gassing of the epoxy used to bond glass substrates, and/or by temperature gradients applied to the optics by the Sun. Both have been intensively studied with ground experiments, which showed that the PSF should be stable over $\sim$10 years. Out-gassing is expected to be strongest in the first years after launch, and we compared observations of Vela X-1 \citep[][; Table \ref{ta:psfobsid}]{Fuerst2014} taken $\sim300$ days apart to look for broadening in the PSF that might suggest contamination or out-gassing had occurred.

We constructed radial profiles for the three observations of Vela~X-1 \citep{Fuerst2014} for the energy bands used above, and compared the radial profiles directly for each radial bin. We subtracted the late-time profiles from the reference profile (T=0, obs. ID 10002007001) in order to see if there was a significant change in the radial profiles. The radial profiles and the differences in the 3--4.5\,keV band are shown in Figure~\ref{fig:PSFtime}. We find there is no significant change in the radial profiles.

\begin{figure*}
\centering
\includegraphics[width=0.32\textwidth]{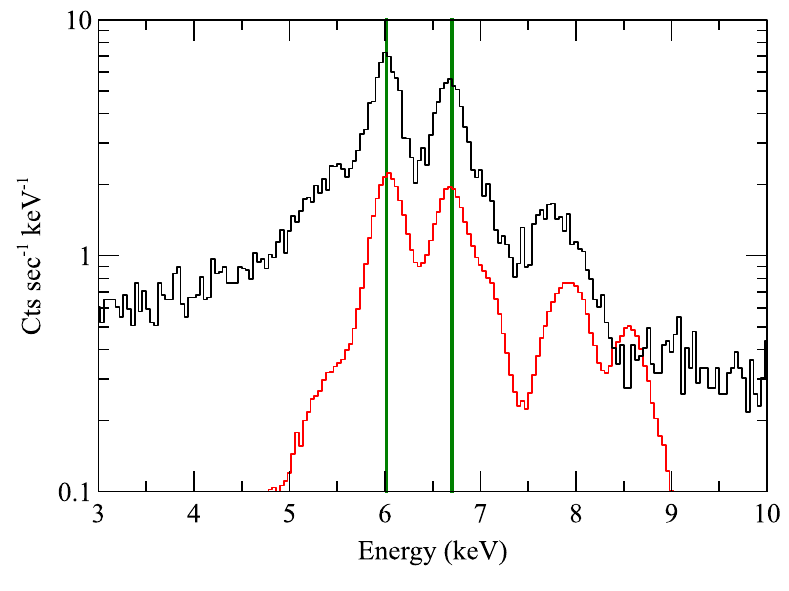}
\includegraphics[width=0.32\textwidth]{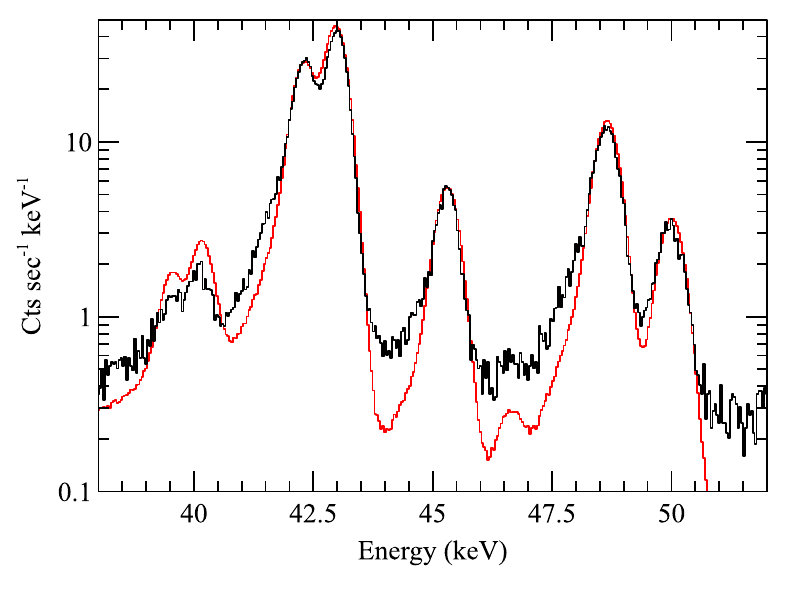}
\includegraphics[width=0.32\textwidth]{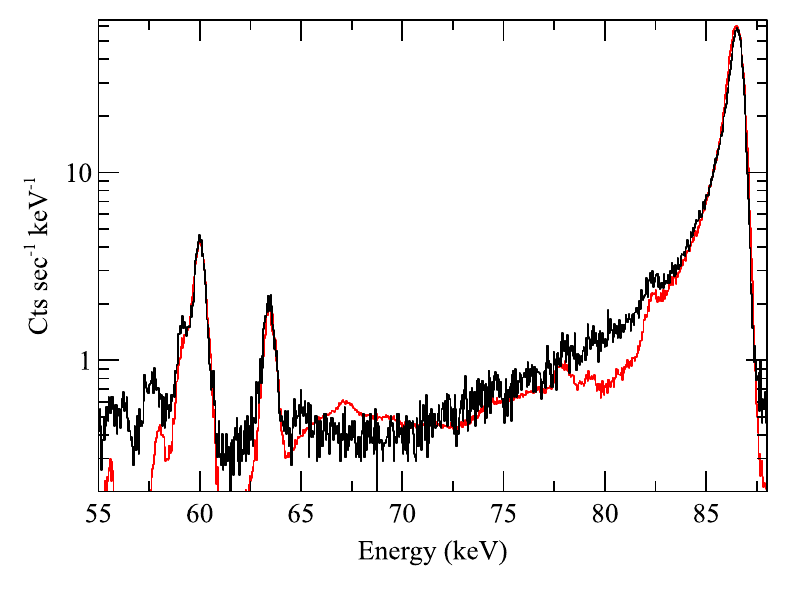}
\caption{Representative spectrum of the calibration source for FPMB Det 0 (black) compared with the spectral model of the calibration source convolved through the instrument response (red, see text). Left: the lowest energy lines along with the locations of the 6.07 and 6.71 keV X-ray lines in the calibration source (vertical green lines.); Middle: The 35 to 55 keV bandpass; Right: the 55 to 88 keV bandpass containing the main 86.54 keV line.}
\label{clc_gain}
\end{figure*}

\begin{figure*}
\centering
\includegraphics[width=0.32\textwidth]{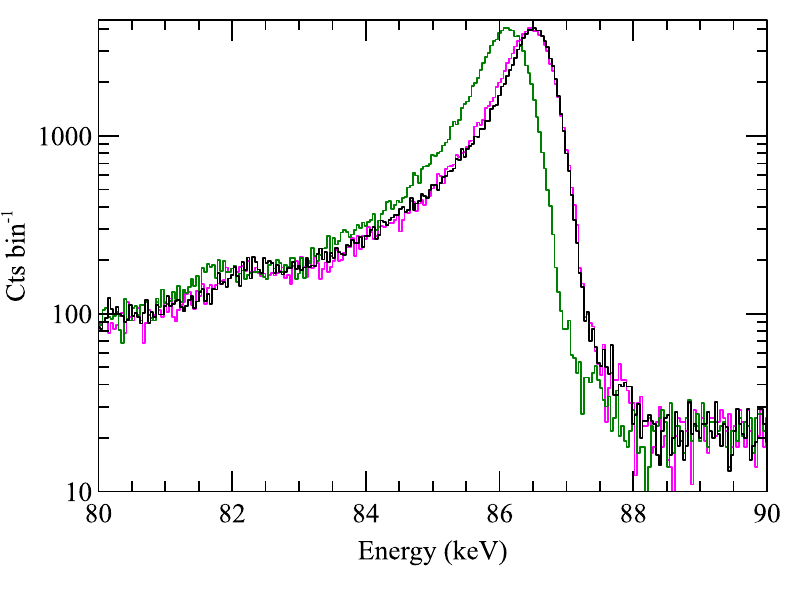}
\includegraphics[width=0.32\textwidth]{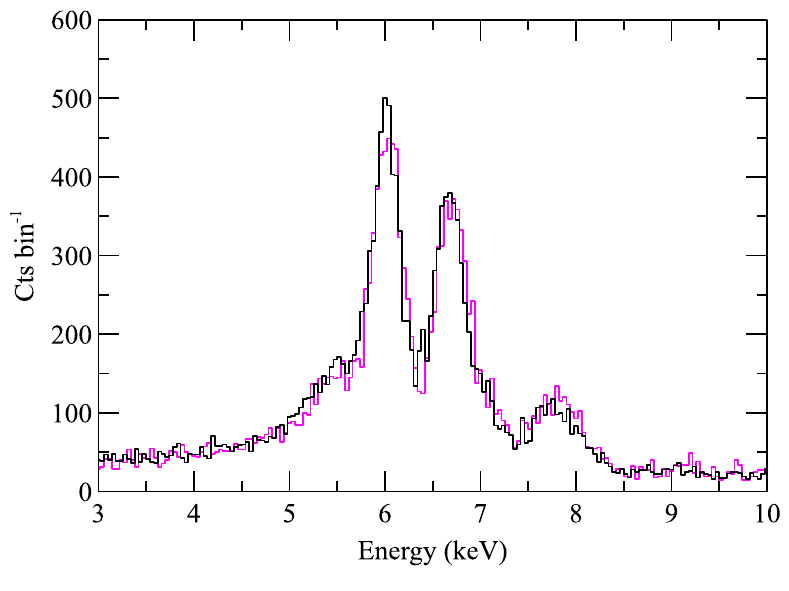}
\includegraphics[width=0.32\textwidth]{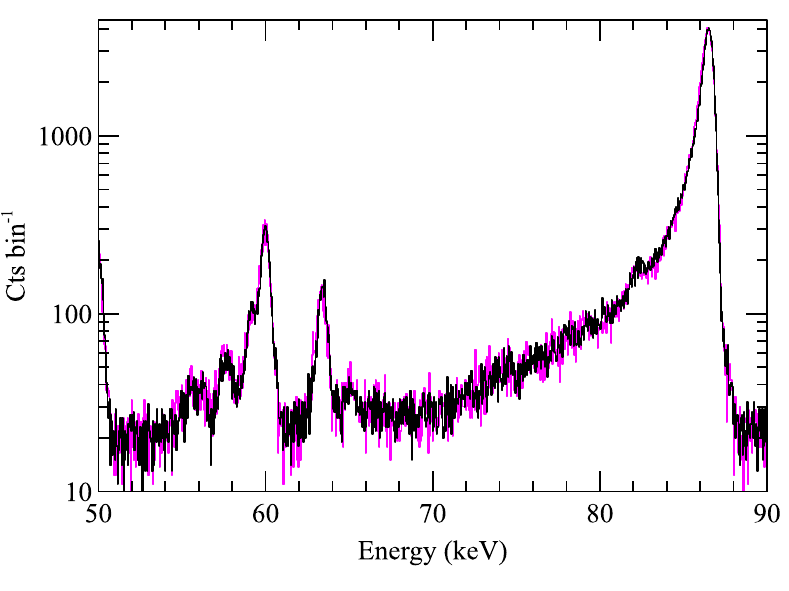}
\caption{Representative spectrum of the calibration source for 2012 epoch FPMB from detector 0 (black) compared with 2015 epoch data from the same detector. Left: A zoomed comparison of the region near the 86.54 keV line show the 2012 data (black) and the 2015 data before (green) and after (magenta) the time-dependent correction has been applied. Middle: A comparison of the 2012 (black) and the corrected 2015 data (magenta) at low-energies; Right: A comparison of the 2012 (black) and corrected 2015 (magenta) data across the 55 to 88 keV bandpass.}
\label{gain_fit}
\end{figure*}

\section{Detector Gain Calibration}\label{gainsection}
Each \nustar\ focal plane contains a calibration source primarily composed of $^{155}$Eu which has several prominent lines at high (86.54 and 105.4\,keV) and low energies (6.06 and 6.71\,keV). The source is located on a movable arm that can deploy the calibration source into the field of view so that we can confirm the gain of the instrument in flight. 

We obtained two epochs of calibration source data: one just after observatory commissioning in June 2012 and one after the prime mission completed in January 2015. The initial values of the CALDB files were determined using extensive pixel-by-pixel-by-grade calibration of the detectors on the ground, where grade refers to the charge sharing pattern among pixels \citep[see][for details]{Kitaguchi2011}. We used the first epoch of calibration data to fine-tune the \texttt{CLC}, ``Charge Loss Coefficient'', and \texttt{FPM\_Gain} CALDB files based on the in-orbit conditions of the instrument. 

As we have limited on-orbit data, we determine the CLC corrections for each detector (e.g. integrated over all pixels) separately for grades 0, 1, 2, 3, and 4 (corresponding to one- and two-pixel events) and as a block for grades 5 through 8 (three-pixel events) and a single correction for grades 9 through 20. We used the second epoch to introduce a time-dependent gain correction in the FPM\_gain CALDB file that takes out an apparent $\approx$0.2\% per year shift in the gain as well as a residual small shift in the zero-point offset of the energy scale (hereafter ``slope" and ``offset", respectively) using the following equation:

\begin{eqnarray}
E_\mathrm{new} &=& \frac{E_\mathrm{old}}{\mathrm{Slope}} - \mathrm{Offset} \,.
\label{eqn:gain}
\end{eqnarray}

We compare the first epoch data to a reference model spectrum to determine the absolute energy scale corrections required (if any) using Equation \ref{eqn:gain}. We constructed a GEANT4 model of the calibration source during ground calibration and produced a simulated spectrum based on the blend of Eu isotopes contained in the calibration source. We multiply this input spectral model by an absorption curve based on the attenuation material between the calibration source and the detectors (primarily the Beryllium entrance windows to the focal plane) and then convolve the result with the detector response matrix (RMF) stored in the CALDB. This results in a ``counts" spectrum that can be directly compared to the observed spectrum to test for changes in the energy scaling. 

We fine-tune the CLC parameters by performing a fit over a change in the slope and the offset and minimizing the resulting chi-square value. The resulting model fits result in large reduced chi-square values ($>5$) owing to the large offset between the model and the observed spectrum at low energies (the result of an unmodeled spectral component likely due to the Compton scattering of gamma-ray photons in the detectors) and small differences in the shape of the 86.54 keV line (only statistically significant due to the large number of source counts near 86 keV). However, the line centroids that we recover are accurate across the \nustar\ science bandpass from 3 to 79 keV (Figure \ref{clc_gain}). 

Iterating this analysis (e.g. applying the correction and then re-fitting the ``new" observed spectrum to the model spectrum) results in residual fits with slope values typically within $2\times10^{-4}$ of unity and offset values within 40 eV of zero. We note that the \nustar\ data are binned to a native resolution of 40 eV, so the fact that we find residual offset errors on that scale is likely evidence for some aliasing. We therefore recommend adopting systematic errors of 40 eV on the offset and $2\times10^{-4}$ on the slope. This implies a systematic uncertainty of 40 eV at energies near Fe emission features in the spectrum and a systematic uncertainty of $\sim$ 60 eV near the 67.86 keV line used for the analysis of $^{44}$Ti (40 eV uncertainty in the offset combined with $\sim$ 20 eV due to the uncertainty in the slope at 68 keV).

Using the second epoch of calibration source data we can determine if there is a long-term trend in the gain of the instrument. We do this by comparing the 2015 epoch data to the 2012 and adjusting the energy channel (PI) value and minimizing the chi-square value between the two distributions. Again, subtle changes in the line shape with time near the 86 keV line and the large number of counts in that line result in a reduced chi-square value that is formally unacceptable (and therefore make it impossible to quote formal statistical errors on the fit parameters). However, we do find by visual inspection that the adjusted 2015 spectra do match the 2012 at all energies (Figure \ref{gain_fit})). 

The corrections required to scale the 2015 up to match the 2012 epoch observations are found in Table \ref{gainfits}. We note that, as above, once these corrections are applied we find residual fits with slopes within $2\times10^{-4}$ of unity and offsets within 40 eV of zero, so we consider these the residual uncertainties on this fitting method. 

The time-dependent gains are implemented in the standard NuSTARDAS pipeline. As of this writing we assume that the time-dependence in linear and can be interpolated between the 2012 and 2015 data sets and can be extrapolated beyond 2015.

\begin{figure*}
\begin{center}
\includegraphics[width=0.23\textwidth]{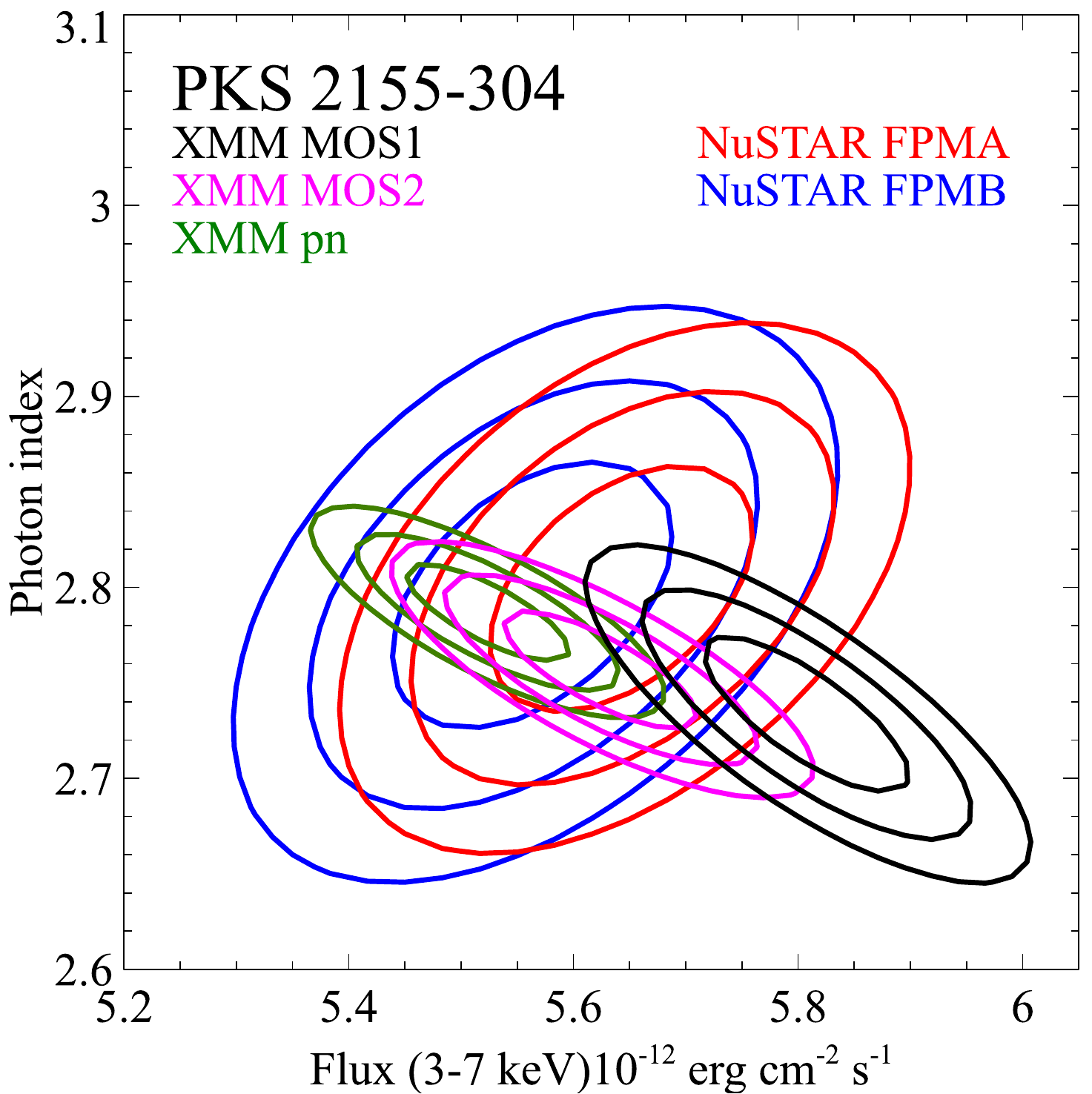}
\includegraphics[width=0.23\textwidth]{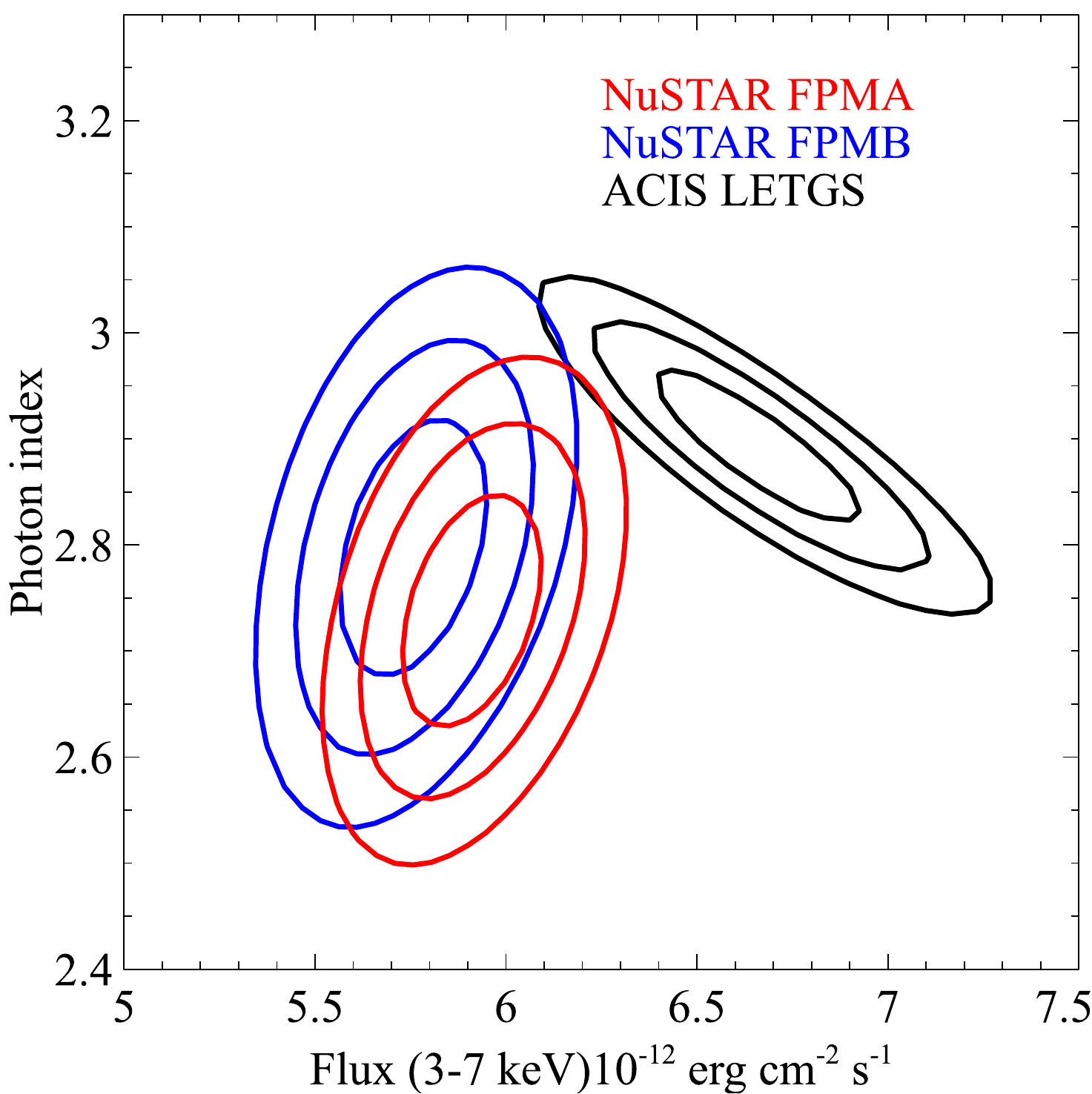}
\includegraphics[width=0.23\textwidth]{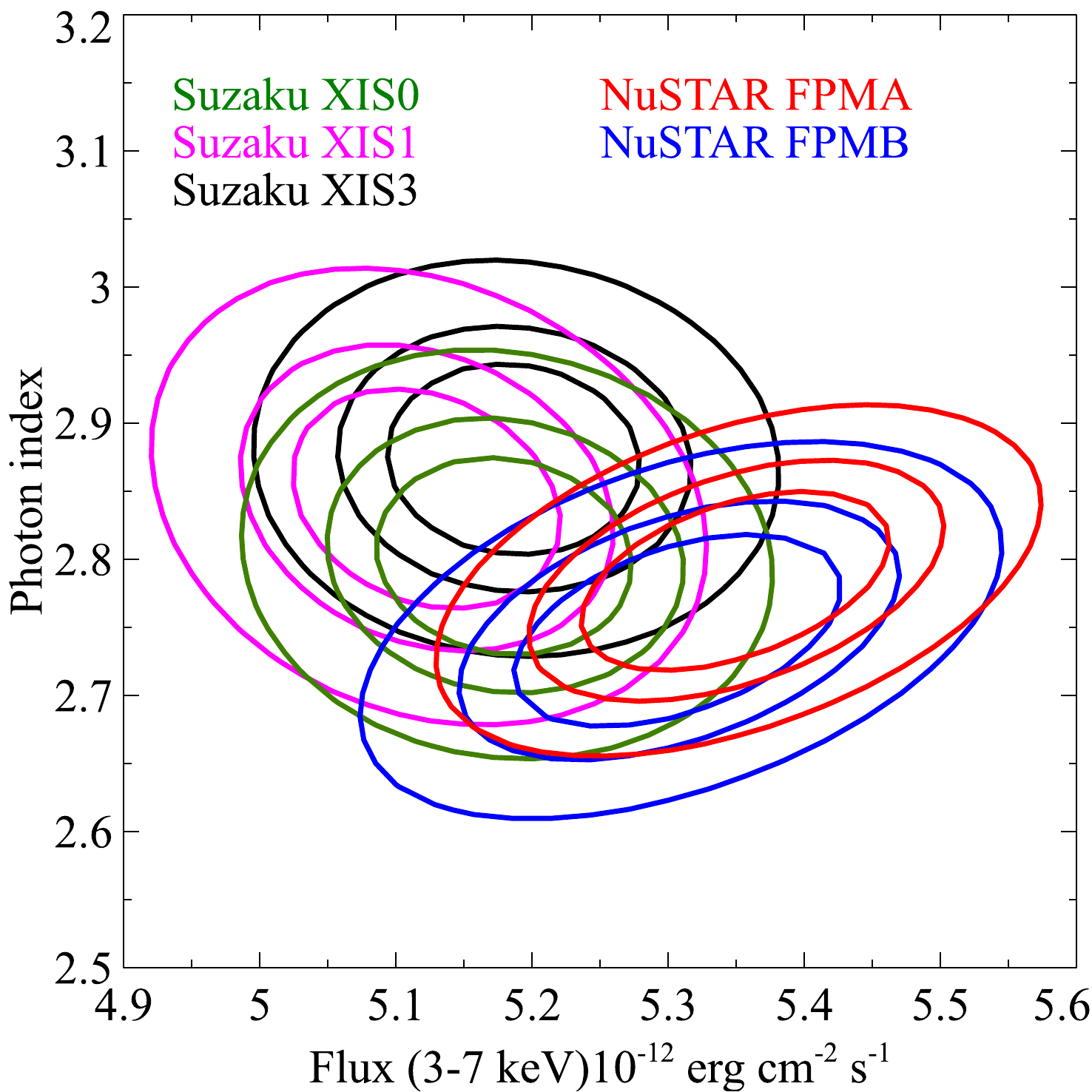}
\includegraphics[width=0.23\textwidth]{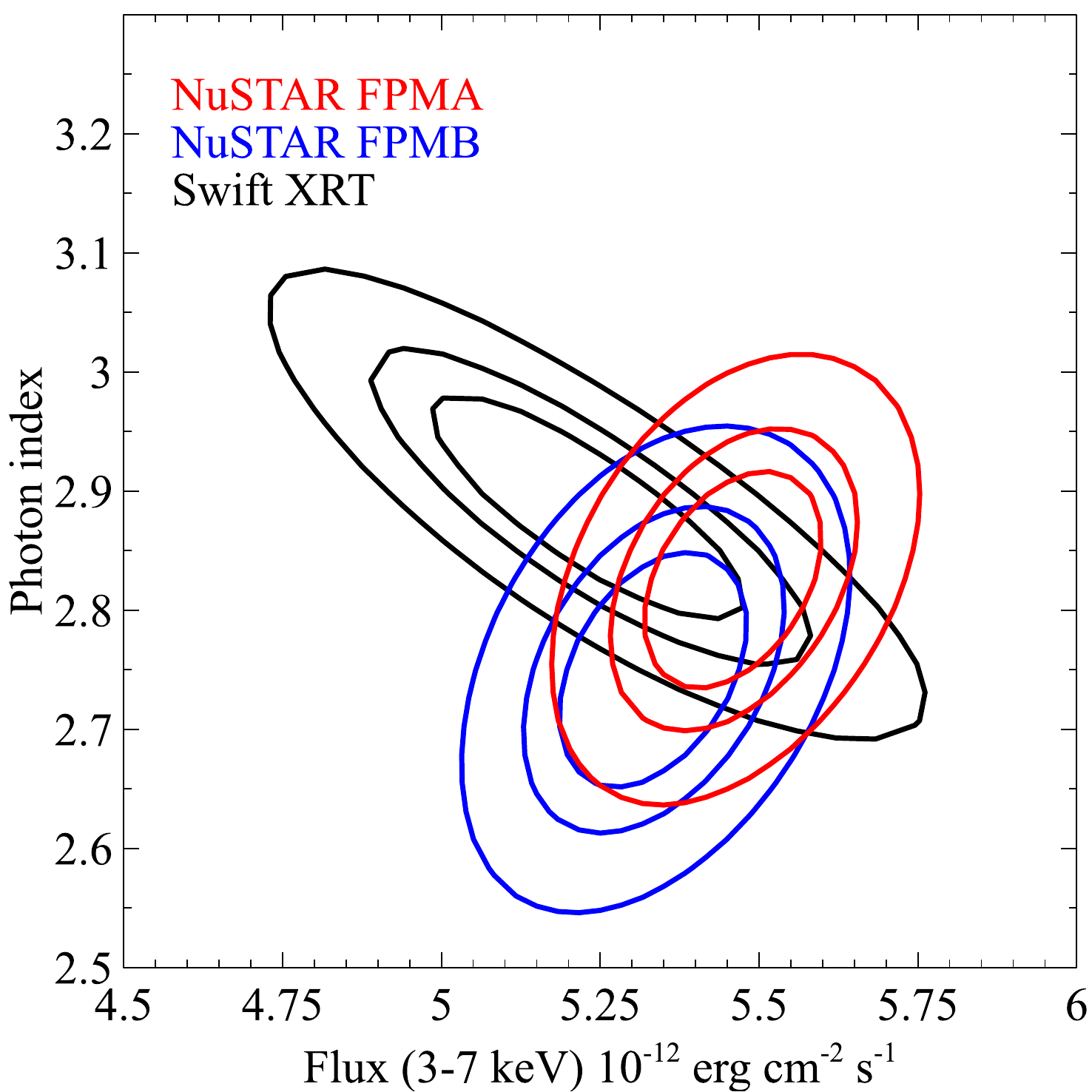}
\includegraphics[width=0.23\textwidth]{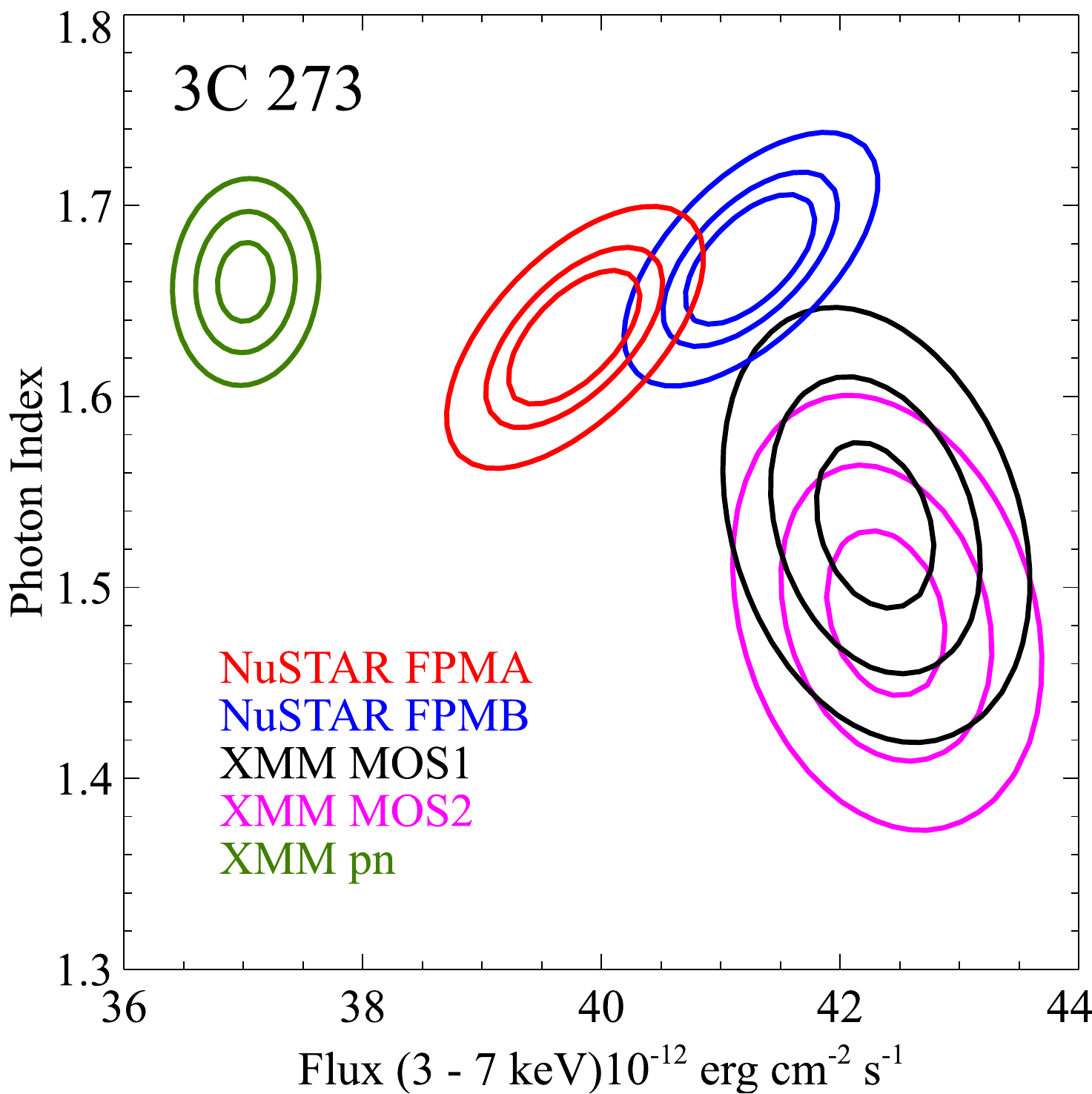}
\includegraphics[width=0.23\textwidth]{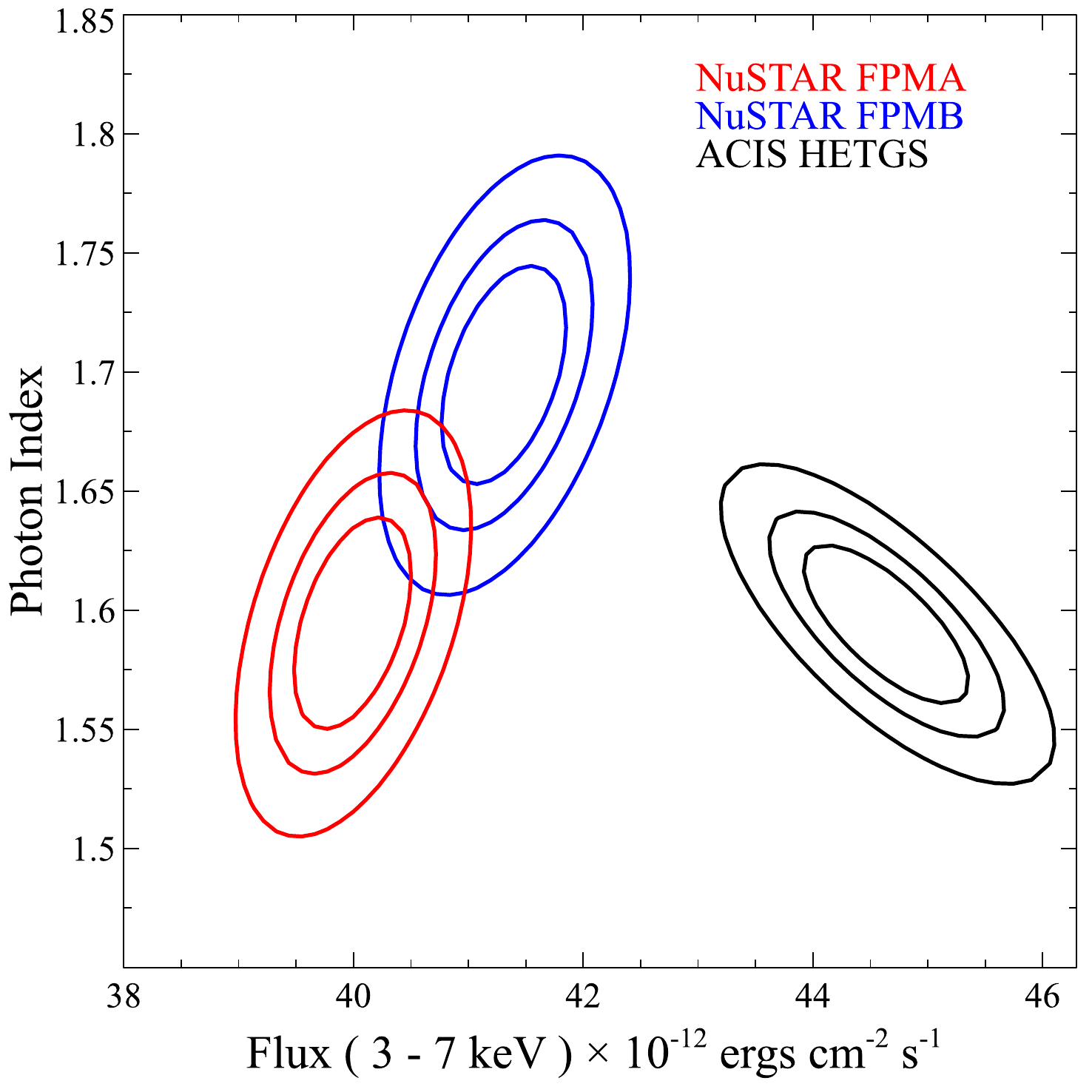}
\includegraphics[width=0.23\textwidth]{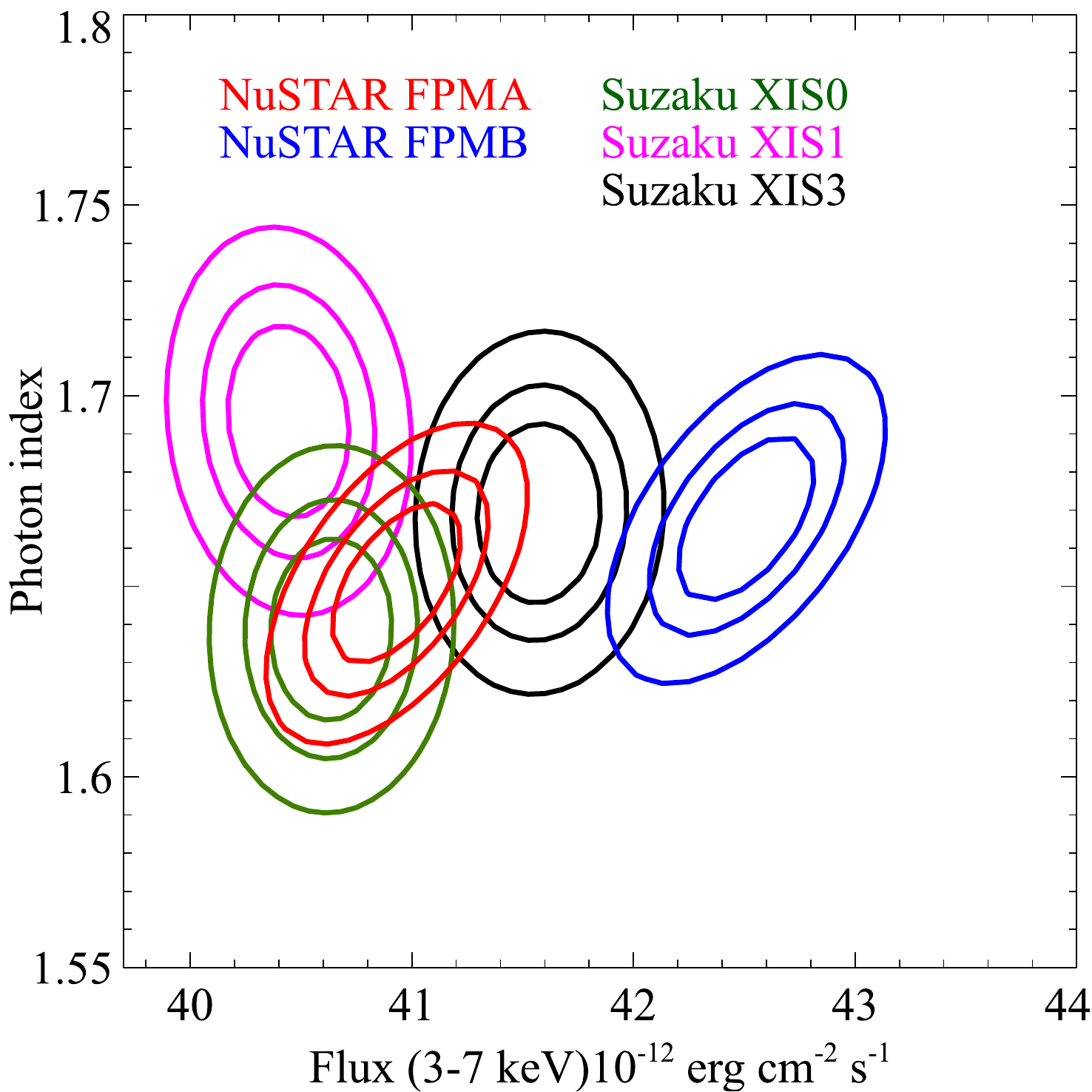}
\includegraphics[width=0.23\textwidth]{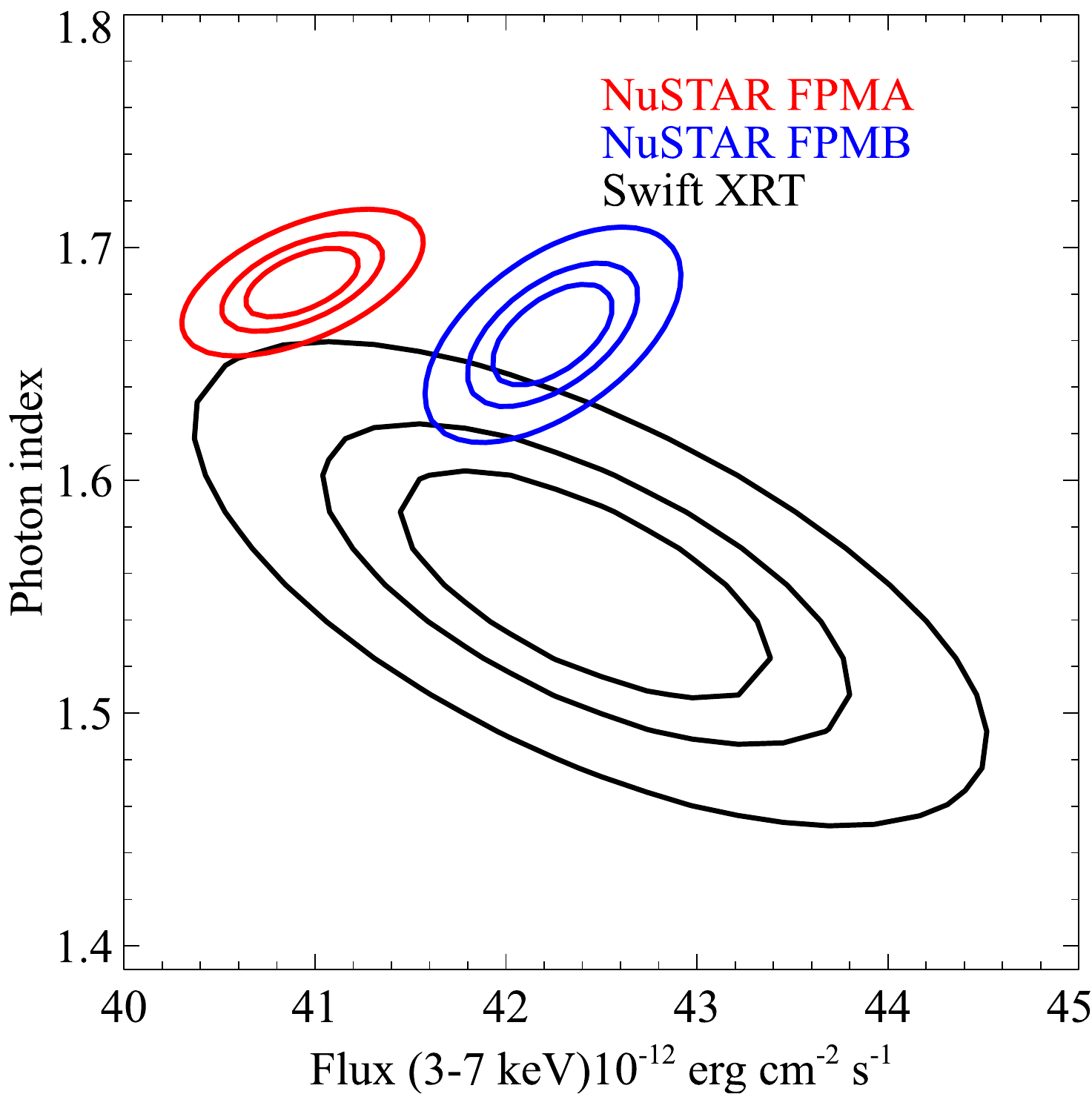}
\includegraphics[width=0.23\textwidth]{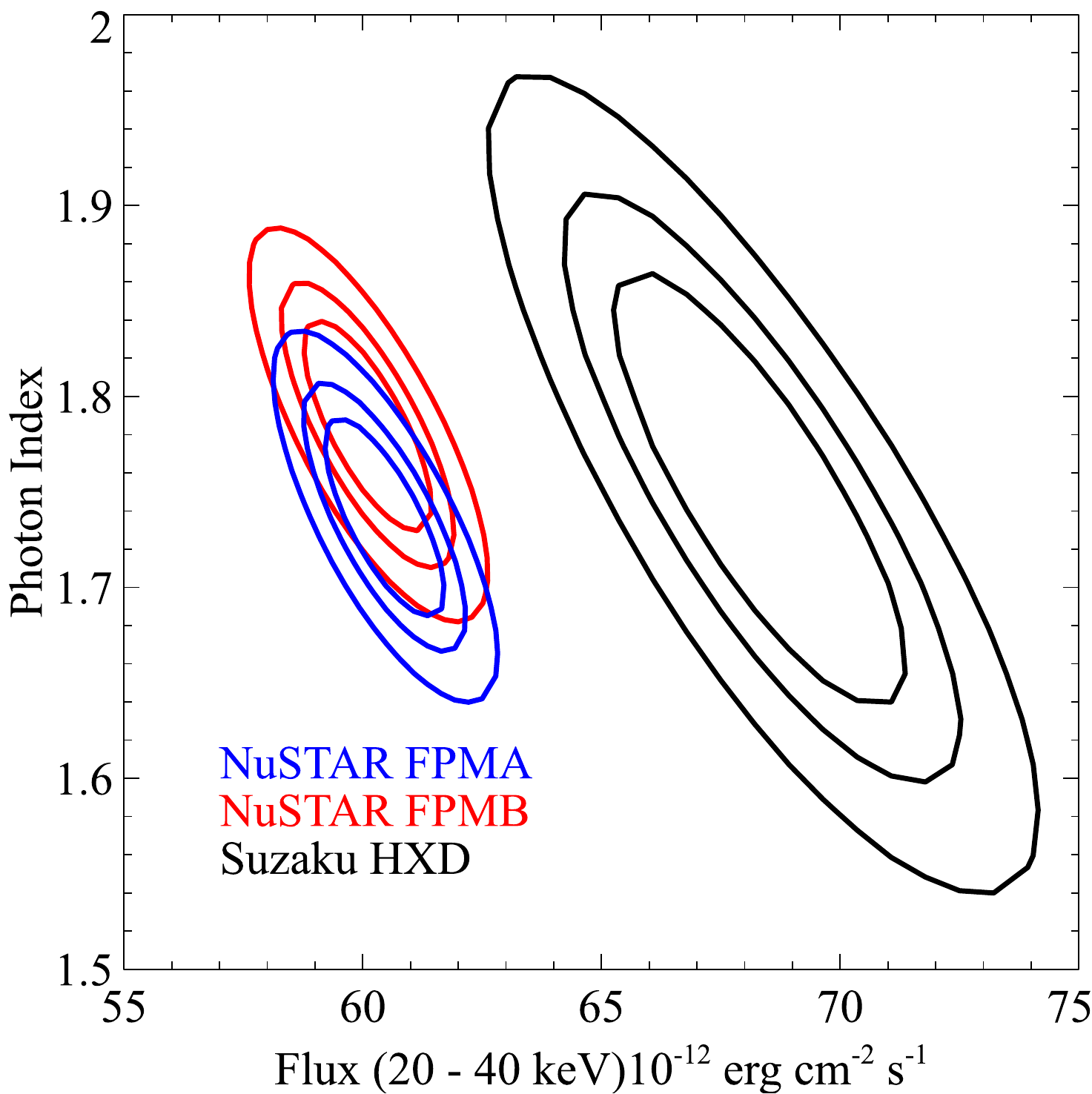}
\end{center}
\caption{Contours of 1, 2 and 3$\sigma$ of the individual fits of each instrument. Top row: PKS\,2155-304. Bottom rows: 3C\,273.}
\label{crossnormcont}
\end{figure*}

\section{Absolute Normalization and Cross-Calibration}\label{crosscalibration}
A set of observations early in the mission, which were individually simultaneous with \textit{Swift}, \textit{Suzaku}, \textit{XMM} and \textit{Chandra}, guided our decision to adjust the final absolute normalization by 15\%, thereby setting the Crab spectrum normalization as measured on average by \nustar\ to N=8.5 keV$^{-1}$ cm$^{-2}$ s$^{-1}$ at 1 keV. This is a flat adjustment at all energies and applied directly into the on-axis effective area files to affect all off-axis angles through the multiplication of the vignetting function. \textit{NuSTAR} participated in the cross-calibration campaign between \textit{Chandra}, \textit{Suzaku}, \textit{Swift} and \textit{XMM-Newton} performed on the quasar 3C\,273 on UT 2012 July 17 and the blazar PKS2155-304 on UT 2013 April 23, and we present here using these two sources the results of the cross-calibration after the normalization correction, addressing the two important questions: (i) How different is \textit{NuSTAR} with respect to the above mentioned observatories? (ii) Accepting the differences, what are the cross-normalization constants that yield the best fit of a model between \textit{NuSTAR} and the other observatories?

Cross-calibration is an ongoing effort coordinated by the International Astronomical Consortium for High Energy Calibration, IACHEC\footnote{http://web.mit.edu/iachec/}, and this analysis is a subset of more detailed forthcoming paper on this campaign. Future cross-calibration campaigns of the IACHEC will expand on and modify these results as the response files of the involved instruments may change.

\subsection{Data Reduction}
The cross-calibration observations are listed in Table \ref{crossobsid} and were processed using the respective standard pipelines.

For \textit{Chandra} we used CIAO 4.6.1 and CALDB 4.6.1.1. For 3C\,273 the data were taken with gratings configuration ACIS+HETG and reprocessed using the CIAO \texttt{chandra\_repro} reprocessing script. We combined orders 1--3 for the HEG and MEG arm separately, and binned the data at 30 counts. For \pksb\ the data were taken with gratings configuration ACIS+LETG. We fit orders m1 and p1 simultaneously. 

For \textit{NuSTAR} we used HEAsoft 6.15.1 and CALDB 20131223. The data were processed with all standard settings and source counts extracted from a 30\as\ radius circular region for both \qcs\ and \pksb. Background was taken from the same detector. We did not combine modules.

For \textit{Suzaku} we used CALDB: HXD (20110913), XIS (20140203). For \qcs\ the observation was taken in 1/4 window mode, and we used 100 arcsec radius circular regions for the front-illuminates, FI, detectors (XIS0,3) and a 140 arcsec region for the back-illuminated, BI, detector (XIS1), such that the regions were centered on the source, but were restricted to the operational portions of the detectors. For \pksb\ the observation was also taken in 1/4 window mode. We extracted from 115\as\ radius circular region for the FI and 135\as\ from the BI. 

For \textit{Swift} we used HEAsoft 6.15.1 and XRT CALDB 2014-02-04. The data for \qcs\ were taken in `PHOTON' mode and \pksb\ in `WINDOW TIMING' mode. Both were reduced using \texttt{xrtpipeline}. Spectra for \qcs\ were extracted from an annulus region, inner radius 5\as\ and outer radius 30\as\ to correct for pileup. The two observations for \qcs\ were combined.

For \textit{XMM-Newton} we used SAS v. 13.5.0 with CALDB 2014-01-31. The data for \qcs\ were taken in `Small Window' mode, and to corrected for pileup in the MOS we extracted counts form an annulus region with inner radius of 15\as\ and outer radius of 45\as\ . For the PN we extracted from a circular region of radius 45\as\ . For \pksb\ the observation was also taken in `Small Window' mode, and for all instruments we extracted from an annular region of inner radius 10\as\ and outer radius 36\as. 

Due to the relative beating of the South Atlantic Anomaly (SAA) passages and occultation periods between the low Earth orbit observatories (\textit{NuSTAR}, \textit{Suzaku}, and \textit{Swift}), we decided to forego strict simultaneity between the observatories. Instead we compare \textit{NuSTAR} exclusively to each instrument separately and truncate the observations to have matching START and STOP times, ignoring SAA passages and occultation periods. 

For the data analysis we use the XSPEC version 12.8.2 analysis software \citep{Arnaud1996}, fitting with \texttt{cstat} statistics \citep{Cash1979} and present $1\sigma$ errors unless otherwise stated.

\subsection{Cross-calibration results}
We fitted the spectra independently in XSPEC with an absorbed pegged power-law (\texttt{tbabs}$\times$\texttt{pegpwrlw}) since it uses as normalization the flux between two energies and breaks the coupling of the regular model between power-law index and normalization at 1\,keV. In this manner we can distinguish between slope differences and overall normalization offsets. The hydrogen column was held fixed in all cases at 1.79 $\times 10^{20}$ cm$^{-2}$ for 3C\,273 and 1.42 $\times 10^{20}$ cm$^{-2}$ for PKS2155-304 \citep{Dickey1990}. The flux normalization of the \texttt{pegpwrlw} was calculated between 3--7\,keV for the soft instruments and 20--40\,keV for the hard instruments. We limited the \textit{NuSTAR} fitting range to match that of the comparison instrument and set the lower energy for the soft instruments to 3\,keV, except for \textit{Chandra} and \textit{Swift} where it was necessary to go down to 2\,keV. While the non-overlapping energy bands may affect the slope measurement, it does not affect the flux calculation. The upper range was 8\,keV for \textit{Chandra}, 8\,keV for \textit{Swift}, 9\,keV for \textit{XMM-Newton} and 9\,keV for \textit{Suzaku}. 

The results are summarized in Table \ref{crossnorm1} for both sources, and Figure \ref{crossnormcont} shows the 1, 2 and 3$\sigma$ contours of the individual fits of the instruments. In all cases there is consistency between the measured slopes when considering the 3$\sigma$ confidence contours, but obvious slope differences do exist, most notably between \nustar\ and \textit{XMM-newton}/MOS instruments and \textit{Swift}/XRT in 3C\,273. It is possible that this discrepancy is in part due to the brightness of 3C\,273, which caused pileup in both instruments. While excision of the piled-up region is common practice, the errors in the PSF wings of the instruments start to play a role and may have skewed the results.

The ratio of fluxes are shown relative to FPMA and \textit{NuSTAR} was intentionally calibrated to be roughly in the middle of the spread between instruments with \textit{Chandra} gratings and \textit{Suzaku}/HXD yielding the highest fluxes (R$_\mathrm{FPMA}\sim 1.10$) and \textit{Swift}/XRT generally the lowest fluxes between 3--7\,keV (R$_\mathrm{FPMA}\sim 0.95$). As discussed in \S\,\ref{repeatability} we expect to see differences between FPMA and FPMB due to the restrictions on the absolute optical axis knowledge and while for PKS2155-304 the two modules yielded the same flux, during 3C\,273 there was an offset of 3\%. It is well known that in general \textit{XMM-Newton}/pn is low compared to MOS \citep{Read2014}, but for 3C\,273 the spectrum appears to be unusually low for reasons that are not understood. This and the slope differences will be investigated in \citet{iachec2015}.

To investigate cross-normalization constants, we made simultaneous fits between each instrument respectively and FPMA and FPMB assuming \texttt{constant}$\times$\texttt{tbabs}$\times$\texttt{pegpwrlw}. We tied all parameters together and only allowed the constant between them to float to evaluate a potential skewing the slope differences might cause on the cross-calibration constant compared to the ratio of fluxes. The results are summarized in Table \ref{crossnorm2}. In all cases the constants are very similar to the FPMA normalized flux from Table \ref{crossnorm1}. We do, however, strongly encourage users to first evaluate the parameters of individual fits and calculate fluxes before before applying a cross-normalization constant and tying fit parameters together. Because of the possible variations in flux between the \nustar\ FPMs, we advise applying a floating cross-normalization constant, but with the caveat that if it takes on extreme values the data reductions might have to be re-examined, and if significant residuals occur after fitting it may be an indication that the source signal-to-noise is high enough to make inter-instrumental slope differences significant. Large discrepancies between FPMA and FPMB may be due to the interference of a gap, which will not affect the shape of the spectrum, but cause one module to have a lower flux.

Different cross-normalization constants may occur if fluxes for energy ranges other than those presented here are used. This is due to slope differences between observatories, and if simultaneity between observations is not strictly enforced this may also have an effect. In general, however, typical values found for other sources than those discussed above are within 10\% as demonstrated in \citet{Walton2014} for Holmberg IX X-1 (\textit{XMM-Newton} and \textit{Suzaku}), \citet{Brenneman2014} for IC 4329A (\textit{Suzaku}) and \citet{Balokovic2014} for NGC 424, NGC 1320, and IC 2560 (\textit{XMM-Newton} and \textit{Swift}).

We have in the above not included the $\sim$3\% uncertainty on the \nustar\ fluxes, which are due to the constraints on absolute knowledge of the optical axis location as discussed in Section \ref{repeatability}. Finally we stress that since both targets are variable and observations were of different lengths between instruments, the non-\nustar\ instrument fluxes should not be compared against each other, only with respect to \nustar. A forthcoming paper \citet{iachec2015} of the IACHEC consortium on these same cross-calibration campaigns, will discuss the cross-calibration between all instruments in far greater detail.

\begin{table}
\centering
\caption{Cross-calibration \textit{tbabs} $\times$ \textit{pegpwrlw}}
\begin{tabular}{l|c|c|c}
\hline
Instrument & $\Gamma$\footnote{Individual fits to each instrument pair.} & Flux 3--7 keV & R$_\mathrm{FPMA}$\footnote{Ratio of flux relative to FPMA, Flux$_X$/Flux$_\mathrm{FPMA}$} \\
& & ($10^{-12}$ erg cm$^{-2}$s$^{-1}$) & \\ 
\hline
\multicolumn{4}{c}{PKS\,2155-304}\\
\hline
FPMA &  2.81 $\pm$ 0.13 & 5.9 $\pm$ 0.2 & 1.0 \\
FPMB &  2.77 $\pm$ 0.13 & 5.8 $\pm$ 0.2 & 0.98 $\pm 0.05$\\
ACIS LETGS &  2.86 $\pm 0.10$ & 6.6 $\pm$ 0.3 & 1.12 $\pm 0.06$\\
\hline
FPMA & 2.78 $\pm$ 0.07 & 5.3 $\pm$ 0.1 & 1.0 \\
FPMB & 2.75 $\pm$ 0.07 & 5.3 $\pm$ 0.1 & 1.0 $\pm$ 0.03\\
XIS0 & 2.8 $\pm$ 0.09 & 5.2 $\pm 0.1$ & 0.98 $\pm$ 0.03\\
XIS1 & 2.84 $\pm$ 0.09 & 5.1 $\pm 0.1$ & 0.96 $\pm$ 0.03\\
XIS3 & 2.87 $\pm$ 0.08 & 5.2 $\pm 0.1$ & 0.98 $\pm$ 0.03\\
\hline
FPMA & 2.82 $\pm$ 0.1 & 5.5 $\pm$ 0.1 & 1.0 \\
FPMB & 2.74 $\pm$ 0.1 & 5.3 $\pm$ 0.1 & 0.96 $\pm$ 0.03\\
XRT & 2.88 $\pm$ 0.1 & 5.2 $\pm$ 0.3 & 0.95 $\pm$ 0.06\\
\hline
FPMA & 2.8 $\pm$ 0.08 & 5.6 $\pm$ 0.1 & 1.0 \\
FPMB & 2.8 $\pm$ 0.08 & 5.6 $\pm$ 0.1 & 1.0 $\pm$ 0.03\\
MOS1 & 2.71 $\pm$0.04 & 5.8 $\pm$0.1 & 1.04 $\pm$ 0.03\\
MOS2 & 2.76 $\pm$0.04 & 5.6 $\pm$0.1 & 1.0 $\pm$ 0.03\\
pn & 2.78 $\pm$0.04 & 5.5 $\pm$0.1 & 0.98 $\pm$ 0.03\\
\hline
\multicolumn{4}{c}{3C\,273}\\
\hline
FPMA &  1.52 $\pm$ 0.09 & 40.3 $\pm$ 0.6 & 1.0 \\
FPMB &  1.68 $\pm$ 0.09 & 41.3 $\pm$ 0.7 & 1.02 $\pm 0.02$\\
ACIS HEGTS &  1.52 $\pm 0.05$ & 44.6 $\pm$ 1.0 & 1.10 $\pm$ 0.03\\
\hline
FPMA & 1.66 $\pm$ 0.02 & 40.9 $\pm$ 0.3 & 1.0 \\
FPMB & 1.66 $\pm$ 0.02 & 42.5 $\pm$ 0.3 & 1.04 $\pm$ 0.01\\
XIS0 & 1.64 $\pm$ 0.03 & 40.6 $\pm$ 0.3 & 0.99 $\pm$ 0.01\\
XIS1 & 1.69 $\pm$ 0.03 & 40.4 $\pm$ 0.3 & 0.99 $\pm$ 0.01\\
XIS3 & 1.66 $\pm$ 0.03 & 41.6 $\pm$ 0.3 & 1.02 $\pm$ 0.01\\
\hline
FPMA & 1.65 $\pm$ 0.03 & 40.9 $\pm$ 0.4 & 1.0 \\
FPMB & 1.66 $\pm$ 0.03 & 42.4 $\pm$ 0.4 & 1.04 $\pm$ 0.01\\
XRT & 1.55 $\pm$ 0.06 & 42.4 $\pm$ 1.2 & 1.04 $\pm$ 0.03 \\
\hline
FPMA & 1.63 $\pm$ 0.04 & 39.8 $\pm$ 0.6 & 1.0 \\
FPMB & 1.68 $\pm$ 0.04 & 41.3 $\pm$ 0.6 & 1.04 $\pm$ 0.02\\
MOS1 & 1.47 $\pm$0.04 & 40.7 $\pm$0.4 & 1.02 $\pm$ 0.02\\
MOS2 & 1.46 $\pm$0.04 & 39.0 $\pm$0.4 & 0.98 $\pm$ 0.02\\
pn & 1.59 $\pm$0.02 & 37.0 $\pm$0.2 & 0.93 $\pm$ 0.01\\
\hline
\multicolumn{4}{c}{}\\
\hline
Instrument & $\Gamma$ & Flux 20--40 keV & Relative \\
& & ($10^{-12}$ erg cm$^{-2}$s$^{-1}$) & Flux \\ 
\hline
FPMA & 1.73 $\pm$ 0.06 & 60.4 $\pm$ 1.4 & 1.0 \\
FPMB & 1.83 $\pm$ 0.06 & 60.6 $\pm$ 1.4 & 1.0 $\pm$ 0.03\\
HXD & 1.75 $\pm$ 0.02 & 68 $\pm 3$ & 1.12 $\pm$ 0.03\\
\hline
\end{tabular}
\label{crossnorm1}
\end{table}

\begin{table}
\centering
\caption{Cross-calibration \textit{constant} $\times$ \textit{tbabs} $\times$ \textit{pegpwrlw}}
\begin{tabular}{l|c|c|c}
\hline
Instrument & $\Gamma$\footnote{Simultaneous fits of each instrument with \nustar, only Constant allowed to float} & Flux 3--7 keV & Constant\footnote{Constant$_\mathrm{FPMA}$ = 1.} \\
& & ($10^{-12}$ erg cm$^{-2}$s$^{-1}$) & \\ 
\hline
\multicolumn{4}{c}{PKS\,2155-304}\\
\hline
ACIS LETGS & 2.83 $\pm 0.06$ & 5.9 $\pm$ 0.2 & 1.15 $\pm$ 0.06\\
XIS0 & 2.80 $\pm$ 0.05 & 5.5 $\pm 0.1$ & 0.95 $\pm$0.03\\
XIS1 & 2.81 $\pm$ 0.05 & 5.5 $\pm 0.1$ & 0.93 $\pm$0.03\\
XIS3 & 2.82 $\pm$ 0.04 & 5.5 $\pm 0.1$ & 0.94 $\pm$0.03\\
XRT & 2.83 $\pm$ 0.05 & 5.4 $\pm$ 0.2 & 1.00 $\pm$0.05 \\
MOS1 & 2.76 $\pm$0.02 & 5.6 $\pm$0.1 & 1.03 $\pm$0.03\\
MOS2 & 2.77 $\pm$0.02 & 5.6 $\pm$0.1 & 1.00 $\pm$0.03\\
pn & 2.74 $\pm$0.02 & 5.6 $\pm$0.1 & 0.98 $\pm$0.03\\
\hline
\multicolumn{4}{c}{3C\,273}\\
\hline
ACIS HETGS & 1.57 $\pm 0.05$ & 40.0 $\pm$ 0.6  & 1.10 $\pm$ 0.04\\
XIS0 & 1.65 $\pm$ 0.01 & 40.9 $\pm 0.3$ & 0.99 $\pm$0.01\\
XIS1 & 1.67 $\pm$ 0.01 & 41.1 $\pm 0.3$ & 0.99 $\pm$0.01\\
XIS3 & 1.66 $\pm$ 0.01 & 41.0 $\pm 0.3$ & 0.01 $\pm$0.01\\
XRT & 1.63 $\pm$ 0.03 & 40.3 $\pm$ 0.5 & 0.99 $\pm$0.03\\
MOS1 & 1.63 $\pm$0.03 & 39 $\pm$0.5 & 1.05 $\pm$0.02\\
MOS2 & 1.62 $\pm$0.02 & 39.7 $\pm$0.5 & 1.06 $\pm$0.02\\
pn & 1.65 $\pm$0.02 & 40.0 $\pm$0.5 & 0.93 $\pm$0.01\\
\hline
\multicolumn{4}{c}{}\\
\hline
Instrument & $\Gamma$ & Flux 20--40 keV & Constant \\
& & ($10^{-12}$ erg cm$^{-2}$s$^{-1}$) & \\ 
\hline
HXD & 1.77 $\pm$ 0.04 & 60 $\pm 1$ & 1.13 $\pm$0.04\\
\hline
\end{tabular}
\label{crossnorm2}
\end{table}

\section{Timing Calibration}\label{time}

\subsection{Estimated Performance}
A basic description of the \textit{NuSTAR} timing system was provided in \cite{Harrison2013} and \cite{Bachetti2015}. We elaborate upon the system here and describe efforts to characterize the timing performance. The performance of the components is summarized in Table \ref{ta:timing}, and described in more detail below.  Known biases in the system are shown in one column, and are typically removed by the standard analysis.  The unmodeled terms are shown in the second column, and are either unknown at the time of writing, or are stochastic.  Based on this analysis, we expect the corrected event time stamps to be accurate at the 3 ms level (95\%).

The \textit{NuSTAR} on-board time reference is a temperature-compensated
crystal oscillator with a nominal frequency of 24\,MHz. The
frequency stability of this oscillator is specified to be $\pm$4\,ppm
over the full operating temperature range. On-board time synchronization of
subsystems is done by broadcasting a one pulse-per-second (PPS)
electronic signal. The propagation delay of this signal is expected to be less than a few microseconds.
The instrument focal plane modules maintain their own internal high
resolution timers: seconds are marked using the PPS signal, and
sub-seconds are marked internally with 2\,$\mu$s precision. The delay between when a photon interacts
with the detector and when it is time-stamped by the electronics is
expected to be a few microseconds, but is certainly less than 0.1\,msec.
After the spacecraft clock timestamp is applied to an event, it is
telemetered to the ground. Science event data is archived with this
timestamp.

The spacecraft clock is routinely correlated with ground time
references during most ground contacts, which allows measurement of the 
spacecraft clock offset and trends.  The offset is measured
essentially by performing one-way ranging of telemetry frames set 1
second intervals.  The time of transmission as recorded on the
spacecraft and the time of reception on the ground as recorded by the
ground station are used to estimate the clock offset.  During downlink
contacts, the rising edge of the PPS signal initiates the downlink of
a telemetry frame.  There is a delay of at least
352.75\,ms before this packet is transmitted to the ground.  During
times when the downlink system is under heavy load, this delay can be
larger, so clock offset data are filtered to consider only idle
downlink times.  One-way range delays are modeled using the known
spacecraft ephemeris, which is typically accurate to $\sim$30 km
($\sim$100 lt-$\mu$s).  The ground stations maintain their own time
standards referenced to UTC via GPS, which are typically accurate at
the few nanosecond level.  However, USN Hawaii and KSAT Singapore
stations only record receive times with 1\,ms precision.  The RF
propagation and Malindi ground station portions of the system have
been validated by analysis of \textit{Swift} data to better than 0.1\,ms
\citep{Cusumano2012}.  

The clock offset trend is examined on the ground. The \textit{NuSTAR} system is required to maintain on-board time to within
100\,ms of UTC, but typically the offset is maintained to less than
30\,ms.  As the spacecraft clock's enviroment changes, the true
frequency also changes. Mission operations staff can adjust the clock
divisor setting from the nominal 24 MHz value in order to steer the
clock offset towards zero.  The clock divisor is the ``effective''
frequency, or oscillator ticks required to produce a PPS pulse.  The
mission-average clock divisor offset is $\sim$14 ppm, with a variation
of about $\pm$0.2 ppm (95\%).  Clock divisor adjustments are visible
in both the science event timestamps and the clock offset
measurements, and are included in the clock modeling as described
below.

We model the clock offset trend as a piece-wise continuous polynomial function.  Control points are inserted
at points in time when the clock divisor was adjusted, and also where
residuals demand extra compliance.  The coefficients of the function
are recorded in a clock correction file, which is available to the
\textit{NuSTAR} community via the CALDB system.  After modeling, the
residuals are less than 2\,ms (95\%), although in cases where the clock
measurements are sparse, the retrieval error may be larger than this.
This file can be used by the general-purpose FTOOL {\tt barycorr} in
order to correct event times to the solar system barycenter time
system, or TDB.  As the \textit{NuSTAR} orbit is known to
approximately 30\,km, light time errors would be less than 0.1\,ms.

\begin{table}
\centering
\caption{\textit{NuSTAR} Expected Timing Performance\label{ta:timing}}
\begin{tabular}{lrr}
\hline
Error Term & Modeled Bias & Unmodeled \\ 
           & (ms)         &  (ms)     \\
\hline
\hline
Spacecraft 1 PPS time transfer &    --- & $<$0.1 \\
Instrument time stamp          &    --- & $<$0.1 \\
Polynomial model               & $<$100 & $<$2   \\
Spacecraft 1 PPS downlink delay& 352.75 & $<$1 \footnote{During idle downlink periods}\\
Range delay                    & $<$10  & $<$0.1 \\
Ground station time stamp      &    --- & $<$0.1\footnote{Malindi ground station}\\
                               &    --- & $<$1\footnote{USN Hawaii ground station}\\
\hline
\hline
\end{tabular}
\end{table}

\begin{figure}
\includegraphics[width=\columnwidth]{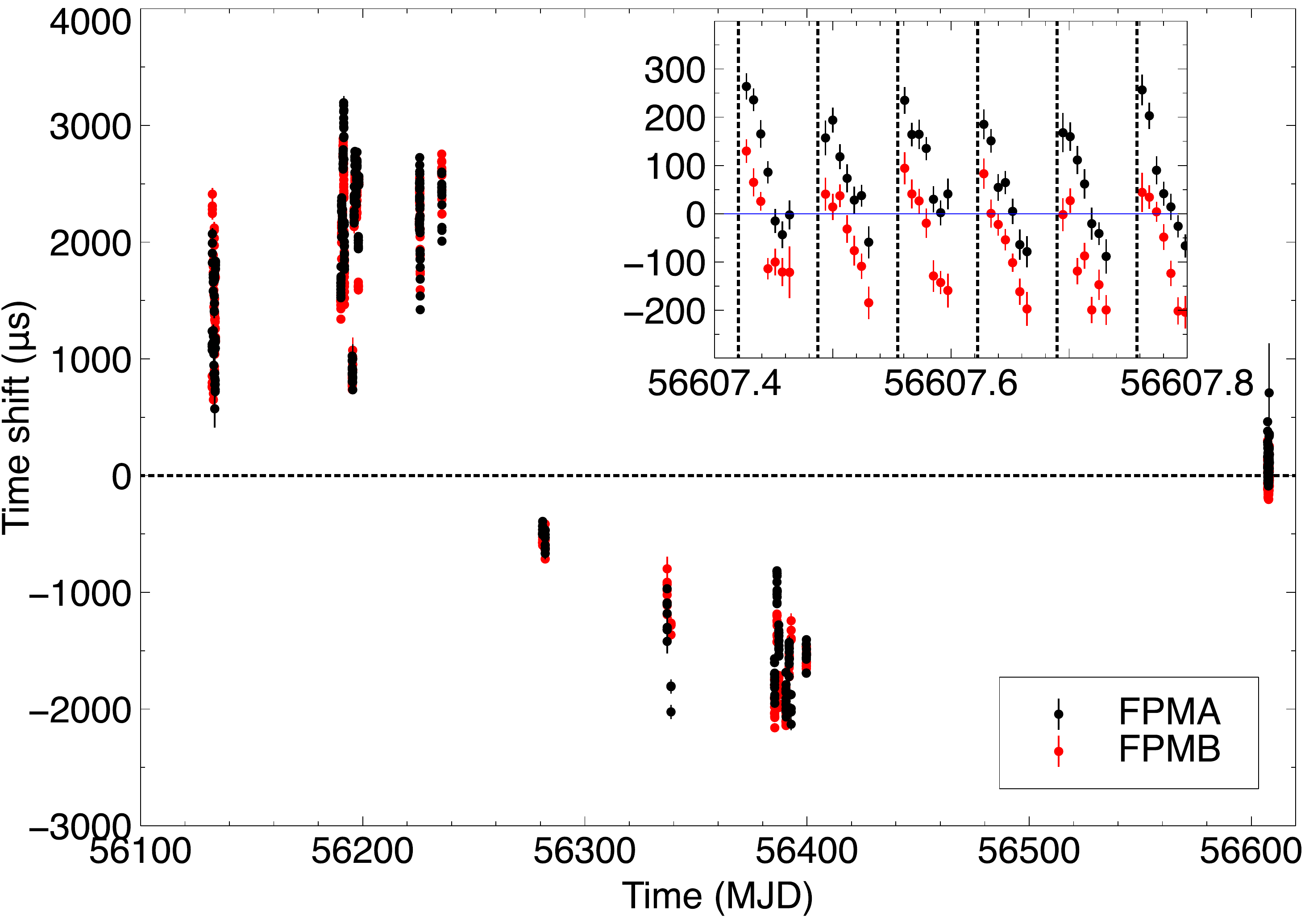}
\caption{Comparison of the TOAs of the Crab and the expected arrival times calculated through the Jodrell Bank Monthly Ephemeris in all available {\em NuSTAR} observations. The historical residuals are about $\pm 3$\,ms. The inset shows a detail of the October 2013 observation. This plot shows a clear shift by about $0.4\pm0.1$\,ms of the measured TOAs, originated by clock frequency variations, on orbital timescales. Dashed lines indicate the orbital timescale for visual purposes.}
\label{crabtime}
\end{figure}

\subsection{Measured Performance}

We used observations of known X-ray pulsars in order to characterize
the actual performance of the system.  This includes absolute timing,
as well as long term stability of the timing measurement system.  We
analyzed the extensive observations of the Crab pulsar described above.
Observation of a single pulsar allows one to track the mission-long
stability of the clock, modulo the $\sim$33\,ms pulse period.  We also
observed PSR B1509$-$58, with a longer period of 151\,ms to resolve
some of this ambiguity.  We describe the results of these observations
below. We refer observations of known pulsars to their absolute radio
ephemerides.

We used all available \textit{NuSTAR} observations of the Crab pulsar (Table \ref{calobsid}), and extracted pulse profiles in the 3--79\,keV range. We used the PRESTO\footnote{\href{http://www.cv.nrao.edu/~sransom/presto/}{http://www.cv.nrao.edu/~sransom/presto/}} \citep{Ransom2001} suite of programs to obtain an average pulsed profile and to calculate the time-of-arrival (TOA) of the pulses on time scales of $\sim5$ minutes. We then compared these TOAs to the relevant monthly timing solution for the Crab produced by the Jodrell Bank radio telescope\footnote{\href{http://www.jb.man.ac.uk/pulsar/crab.html}{http://www.jb.man.ac.uk/pulsar/crab.html}} \citep{Lyne1993} using the Tempo2 software\footnote{\href{http://www.atnf.csiro.au/research/pulsar/tempo2/}{http://www.atnf.csiro.au/research/pulsar/tempo2/}} \citep{Edwards2006}.  The radio ephemeris is referred to well-known terrestrial time standards.  There is an offset between radio and X-ray pulse profiles of $\sim$300\,$\mu$s, which we do not consider further in this analysis because it is negligible in light of the current \textit{NuSTAR} timing performance.

The results are shown in Figure~\ref{crabtime}.  Over the whole
history of \textit{NuSTAR} Crab observations, the residuals of these
TOAs show a trend between $-$2\,ms and $+$3\,ms.  This level
of residual is indeed mostly inside the $\pm\sim3$\,ms uncertainty
range estimated above.  

In addition to the long term residual trend, shorter term trends are
also visible (see inset of Figure~\ref{crabtime}).  These trends
repeat on the time scale of the spacecraft orbital period of about
5400\,s (0.07 day), and have an amplitude of about 400\,$\mu$s.  This
trend is clearly related to the temperature history of the spacecraft
clock oscillator.  Oscillator instability on these time scales hinders
the detection of pulsars with period shorter than 3\,ms, because the
pulse peak will be smeared over a spacecraft orbital period.

We also measured the TOAs of another pulsar, PSR B1509$-$58, with a
pulse period of $\sim151$\,ms and verified that the TOAs were aligned
with those measured by \textit{Swift} in SW timing mode to $1\pm2$\,ms (details
can be found in \citet{Mori2014}.

In summary, using two pulsars with known absolute ephemerides, the
\textit{NuSTAR} timing system is performing at the accuracy level of
3\,ms over the entire mission to date, which is about the expected
level of timing uncertainty based on design (Table~\ref{ta:timing}).

Present research indicates that the spacecraft clock frequency
variations are driven almost entirely by its thermal environment.  As
accurate temperature information about the environment is known, it
may be possible to make a more accurate clock variation model.  This
will be the subject of future work.

\section{Summary}
The \textit{NuSTAR} observatory is well understood, calibrated, and operating within requirements. We have calibrated the instrument effective area responses against the phase-averaged Crab nebula + pulsar out to 7\am\ assuming a canonical spectrum of $\Gamma=2.1$ and normalization $N=8.5$. The scatter in fit values of the Crab data-set after adjustment are summarized in Table \ref{crab_spect_stats} and are a measurement of the repeatability between consecutive observation of the same source with varying off-axis angle ($<3$\am) and position on the detector. For the Crab ensemble we find $\Gamma = 2.1 \pm 0.01$ and normalization $N = 8.5 \pm 0.3$. The flux differences between FPMA and FPMB are of the order 0--5\% and understood to be due to mast motions and the uncertainty in the true optical axis location, which is known to $\sim30$\as. The detectors have between 2012 and 2015 shown a gain shift in slope and offset (summarized in Table \ref{gainfits}) and the estimated uncertainty on the energy is 60\,eV. Cross-calibration campaigns with \textit{Chandra}, \textit{Swift}, \textit{Suzaku} and \textit{XMM-Newton} have shown that the fluxes of the sources \qcs\ and \pksb\  in \textit{NuSTAR} agree with the other observatories to within 10\%. Finally we have demonstrated that \textit{NuSTAR} is capable of 3ms timing accuracy, and we plan to continue improving the timing capability of the observatory. We will continue to monitor the detector gain, PSF and absorption elements for any changes.


{\it Facility:} \facility{CXO}, \facility{NuSTAR}, \facility{Swift}, \facility{Suzaku}, and \facility{XMM}

\acknowledgments
We would like to thank the referee for helpful comments and suggestions which helped improve the paper. This work was supported under NASA Contract No. NNG08FD60C, and made use of data from the NuSTAR mission, a project led by the California Institute of Technology, managed by the Jet Propulsion Laboratory, and funded by the National Aeronautics and Space Administration. We thank the NuSTAR Operations, Software and Calibration teams for support with the execution and analysis of these observations. This research has made use of the NuSTAR Data Analysis Software (NuSTARDAS) jointly developed by the ASI Science Data Center (ASDC, Italy) and the California Institute of Technology (USA).

\bibliography{bib}

\begin{thebibliography}{}

\bibitem[\protect\astroncite{{Arnaud}}{1996}]{Arnaud1996}
{Arnaud} K.A.,  1996,
\newblock In: {Jacoby} G.H., {Barnes} J. (eds.) Astronomical Data Analysis
  Software and Systems V, Vol. 101. Astronomical Society of the Pacific
  Conference Series, p.~17

\bibitem[\protect\astroncite{{Bachetti} et~al.}{2015}]{Bachetti2015}
{Bachetti} M., {Harrison} F.A., {Cook} R., et~al., 2015, \apj 800, 109

\bibitem[\protect\astroncite{{Balokovi{\'c}} et~al.}{2014}]{Balokovic2014}
{Balokovi{\'c}} M., {Comastri} A., {Harrison} F.A., et~al., 2014, \apj 794, 111

\bibitem[\protect\astroncite{{Brejnholt} et~al.}{2011}]{Brejnholt2011}
{Brejnholt} N.F., {Christensen} F.E., {Jakobsen} A.C., et~al., 2011,
\newblock In: Proc. SPIE, Vol. 8147.

\bibitem[\protect\astroncite{{Brejnholt} et~al.}{2012}]{Brejnholt2012}
{Brejnholt} N.F., {Christensen} F.E., {Westergaard} N.J., et~al., 2012,
\newblock In: Proc. SPIE, Vol. 8443.

\bibitem[\protect\astroncite{{Brenneman} et~al.}{2014}]{Brenneman2014}
{Brenneman} L.W., {Madejski} G., {Fuerst} F., et~al., 2014, \apj 788, 61

\bibitem[\protect\astroncite{{Cash}}{1979}]{Cash1979}
{Cash} W.,  1979, \apj 228, 939

\bibitem[\protect\astroncite{{Cirrone} et~al.}{2010}]{Cirrone2010}
{Cirrone} G.A.P., {Cuttone} G., {Di Rosa} F., et~al., 2010, Nuclear Instruments
  and Methods in Physics Research A 618, 315

\bibitem[\protect\astroncite{{Cusumano} et~al.}{2012}]{Cusumano2012}
{Cusumano} G., {La Parola} V., {Capalbi} M., et~al., 2012, \aap 548, A28

\bibitem[\protect\astroncite{{Dickey} \& {Lockman}}{1990}]{Dickey1990}
{Dickey} J.M., {Lockman} F.J.,  1990, \araa 28, 215

\bibitem[\protect\astroncite{{Edwards} et~al.}{2006}]{Edwards2006}
{Edwards} R.T., {Hobbs} G.B., {Manchester} R.N.,  2006, \mnras 372, 1549

\bibitem[\protect\astroncite{{F{\"u}rst} et~al.}{2014}]{Fuerst2014}
{F{\"u}rst} F., {Pottschmidt} K., {Wilms} J., et~al., 2014, \apj 780, 133

\bibitem[\protect\astroncite{{Gehrels} et~al.}{2004}]{Gehrels2004}
{Gehrels} N., {Chincarini} G., {Giommi} P., et~al., 2004, \apj 611, 1005

\bibitem[\protect\astroncite{{Harrison} et~al.}{2013}]{Harrison2013}
{Harrison} F.A., {Craig} W.W., {Christensen} F.E., et~al., 2013, \apj 770, 103

\bibitem[\protect\astroncite{{Ishida} et~al.}{2011}]{Ishida2011}
{Ishida} M., {Tsujimoto} M., {Kohmura} T., et~al., 2011, \pasj 63, 657

\bibitem[\protect\astroncite{{Jansen} et~al.}{2001}]{Jansen2001}
{Jansen} F., {Lumb} D., {Altieri} B., et~al., 2001, \aap 365, L1

\bibitem[\protect\astroncite{{Kirsch} et~al.}{2005}]{Kirsch2005}
{Kirsch} M.G., {Briel} U.G., {Burrows} D., et~al., 2005,
\newblock In: {Siegmund} O.H.W. (ed.) Proc. SPIE, Vol. 5898., p.22

\bibitem[\protect\astroncite{{Kitaguchi} et~al.}{2011}]{Kitaguchi2011}
{Kitaguchi} T., {Grefenstette} B.W., {Harrison} F.A., et~al., 2011,
\newblock In: Proc. SPIE, Vol. 8145.

\bibitem[\protect\astroncite{{Koglin} et~al.}{2011}]{Koglin2011}
{Koglin} J.E., {An} H., {Barri{\`e}re} N., et~al., 2011,
\newblock In: Proc. SPIE, Vol. 8147.

\bibitem[\protect\astroncite{{Kuiper} et~al.}{2001}]{Kuiper2001}
{Kuiper} L., {Hermsen} W., {Cusumano} G., et~al., 2001, \aap 378, 918

\bibitem[\protect\astroncite{{Lyne} et~al.}{1993}]{Lyne1993}
{Lyne} A.G., {Pritchard} R.S., {Graham-Smith} F.,  1993, \mnras 265, 1003

\bibitem[\protect\astroncite{{Madsen}}{prep}]{iachec2015}
{Madsen} K.K.,  in prep

\bibitem[\protect\astroncite{{Madsen} et~al.}{2009}]{Madsen2009}
{Madsen} K.K., {Harrison} F.A., {Mao} P.H., et~al., 2009,
\newblock In: Proc. SPIE, Vol. 7437.

\bibitem[\protect\astroncite{{Mitsuda} et~al.}{2007}]{Mitsuda2007}
{Mitsuda} K., {Bautz} M., {Inoue} H., et~al., 2007, \pasj 59, 1

\bibitem[\protect\astroncite{{Mori} et~al.}{2014}]{Mori2014}
{Mori} K., {Gotthelf} E.V., {Dufour} F., et~al., 2014, \apj 793, 88

\bibitem[\protect\astroncite{{Petre} \& {Serlemitos}}{1985}]{Petre1985}
{Petre} R., {Serlemitos} P.,  1985, Applied Optics 24, 1833

\bibitem[\protect\astroncite{{Rana} et~al.}{2009}]{Rana2009}
{Rana} V.R., {Cook}, III W.R., {Harrison} F.A., et~al., 2009,
\newblock In: Proc. SPIE, Vol. 7435., p.~3

\bibitem[\protect\astroncite{{Ransom}}{2001}]{Ransom2001}
{Ransom} S.M.,  2001,
\newblock PhD thesis, Harvard University

\bibitem[\protect\astroncite{{Read} et~al.}{2014}]{Read2014}
{Read} A.M., {Guainazzi} M., {Sembay} S.,  2014, \aap 564, A75

\bibitem[\protect\astroncite{{Shaposhnikov} et~al.}{2012}]{Shaposhnikov2012}
{Shaposhnikov} N., {Jahoda} K., {Markwardt} C., et~al., 2012, \apj 757, 159

\bibitem[\protect\astroncite{{Tsujimoto} et~al.}{2011}]{Tsujimoto2011}
{Tsujimoto} M., {Guainazzi} M., {Plucinsky} P.P., et~al., 2011, \aap 525, A25

\bibitem[\protect\astroncite{{Verner} et~al.}{1996}]{Verner1996}
{Verner} D.A., {Ferland} G.J., {Korista} K.T., {Yakovlev} D.G.,  1996, \apj
  465, 487

\bibitem[\protect\astroncite{{Walton} et~al.}{2014}]{Walton2014}
{Walton} D.J., {Harrison} F.A., {Grefenstette} B.W., et~al., 2014, \apj 793, 21

\bibitem[\protect\astroncite{{Weisskopf} et~al.}{2002}]{Weisskopf2002}
{Weisskopf} M.C., {Brinkman} B., {Canizares} C., et~al., 2002, \pasp 114, 1

\bibitem[\protect\astroncite{{Weisskopf} et~al.}{2010}]{Weisskopf2010}
{Weisskopf} M.C., {Guainazzi} M., {Jahoda} K., et~al., 2010, \apj 713, 912

\bibitem[\protect\astroncite{{Westergaard} et~al.}{2012}]{Westergaard2012}
{Westergaard} N.J., {Madsen} K.K., {Brejnholt} N.F., et~al., 2012,
\newblock In: Proc. SPIE, Vol. 8443.

\bibitem[\protect\astroncite{{Wik} et~al.}{2014}]{Wik2014}
{Wik} D.R., {Hornstrup} A., {Molendi} S., et~al., 2014, \apj 792, 48

\bibitem[\protect\astroncite{{Wilms} et~al.}{2000}]{Wilms2000}
{Wilms} J., {Allen} A., {McCray} R.,  2000, \apj 542, 914

\bibitem[\protect\astroncite{{Wilson-Hodge} et~al.}{2011}]{Wilson2011}
{Wilson-Hodge} C.A., {Cherry} M.L., {Case} G.L., et~al., 2011, \apjl 727, L40

\bibitem[\protect\astroncite{{Winkler} et~al.}{2003}]{Winkler2003}
{Winkler} C., {Courvoisier} T.J.L., {Di Cocco} G., et~al., 2003, \aap 411, L1

\end{thebibliography}
\bibliographystyle{jwaabib}

\end{document}